\documentclass[%
reprint,
superscriptaddress,
amsmath,amssymb,
mathtools,
aps, 
pra,
]{revtex4-2}

\usepackage{graphicx}
\usepackage[colorlinks=true,allcolors=blue]{hyperref}
\usepackage[separate-uncertainty=true,multi-part-units=single]{siunitx} 

\usepackage{multirow}
\usepackage{booktabs}%

\newcommand{\eref}[1]{Eq.~\ref{#1}}
\newcommand{\fref}[1]{Fig.~\ref{#1}}
\newcommand{\ffref}[1]{Fig.~\ref{#1}}
\newcommand{\tref}[1]{Tab.~\ref{#1}}

\newcommand{\aref}[1]{Suppl.~Mat.~\ref{#1}}

\newcommand{\beginsupplement}{%
        \setcounter{section}{0}
        \setcounter{table}{0}
        \renewcommand{\thetable}{S\arabic{table}}%
        \setcounter{figure}{0}
        \renewcommand{\thefigure}{S\arabic{figure}}%
        \renewcommand{\theHfigure}{S\arabic{figure}}
        \setcounter{equation}{0}
        \renewcommand{\theequation}{S\arabic{equation}}%
                \renewcommand{\thesection}{S\arabic{section}}%
     }

\begin{document}
	
	\title{Superconducting flip-chip devices using indium microspheres on Au-passivated Nb or NbN as under-bump metallization layer}

	\author{Achintya Paradkar}
	\author{Paul Nicaise}
	\author{Karim Dakroury}
	\author{Fabian Resare}
	\author{Witlef Wieczorek}
	\email{witlef.wieczorek@chalmers.se}
	\affiliation{Department of Microtechnology and Nanoscience (MC2),\\Chalmers University of Technology, SE-412 96 Gothenburg, Sweden}
		
	\begin{abstract}
	Superconducting flip-chip interconnects are crucial for the three-dimensional integration of superconducting circuits in sensing and quantum technology applications. We demonstrate a simplified approach for a superconducting flip-chip device using commercially available indium microspheres and an in-house-built transfer stage for bonding two chips patterned with superconducting thin films. We use a gold-passivated niobium or niobium nitride layer as an under-bump metallization (UBM) layer between an aluminum-based superconducting wiring layer and the indium interconnect. At millikelvin temperatures, our flip-chip assembly can transport a supercurrent with tens of milliamperes, limited by the smallest geometric feature size and critical current density of the UBM layer and not by the indium interconnect. We show that the pressed indium interconnect itself can carry a supercurrent exceeding 1\,A due to its large size of about \SI{500}{\micro\meter} diameter. Our flip-chip assembly does not require electroplating nor patterning of indium. The assembly process does not need a flip-chip bonder and can be realized with a transfer stage using a top chip with transparency or through-vias for alignment. These flip-chip devices can be utilized in applications that require few superconducting interconnects carrying large currents at millikelvin temperatures.
	\end{abstract}


	\maketitle

Superconducting flip-chip interconnects are pivotal in the integration of superconducting circuits for quantum computing \cite{Foxen_2017, Rosenberg_2017, Conner_2021, Kosen_2022, Norris_Michaud_Pahl_Kerschbaum_Eichler_Besse_Wallraff_2024} and sensing applications \cite{DeNigris_Chervenak_Bandler_Chang_Costen_Eckart_Ha_Kilbourne_Smith_2018, Lucas_Biesecker_Doriese_Duff_Hilton_Ullom_Vissers_Schmidt_2022}. A commonly used method employs indium (In) bump bonding as In is superconducting below \SI{3.4}{\kelvin}, ductile \cite{Plötner_Sadowski_Rzepka_Blasek_1991}, and mechanically stable at cryogenic temperatures \cite{Foxen_2017, Rosenberg_2017, Lucas_Biesecker_Doriese_Duff_Hilton_Ullom_Vissers_Schmidt_2022}, resilient to thermal cycling \cite{Foxen_2017}, and reduces thermal stress between the top and bottom chips \cite{Plötner_Sadowski_Rzepka_Blasek_1991, Huang_Xu_Yuan_Cheng_Luo_2010}. However, indium as a bonding material may diffuse into certain superconducting thin films, potentially degrading their superconducting properties. This issue arises when indium is bonded to aluminum (Al), which results in an intermetallic compound that exhibits non-superconducting behavior \cite{Wade1973}. Therefore, a commonly used method is to add an under-bump metallization (UBM) layer between the In and Al layers to prevent this diffusion. The UBM layer is usually made with superconductors such as niobium nitride \cite{Gordon_1999, Kosen_2022} or titanium nitride \cite{Foxen_2017, Thomas_Michel_Deschaseaux_Charbonnier_Souil_Vermande_Campo_Farjot_Rodriguez_Romano_et_al._2022}. The UBM layer can potentially have a native oxide that must be removed before indium microbumps are deposited to make galvanic contact.

Indium microbumps are typically grown via electroplating \cite{Huang_Xu_Yuan_Cheng_Luo_2010, Volpert2010IndiumDP} or evaporation \cite{Volpert2010IndiumDP, Foxen_2017, Kosen_2022, Norris_Michaud_Pahl_Kerschbaum_Eichler_Besse_Wallraff_2024}  and face-to-face compression bonded using a flip-chip bonder \cite{Foxen_2017, Kozłowski_Czuba_Chmielewski_Ratajczak_Branas_Korczyc_Regiński_Jasik_2021, Kosen_2022, Das_2019,  Norris_Michaud_Pahl_Kerschbaum_Eichler_Besse_Wallraff_2024}. This technique is used in high-density three-dimensional packaging of integrated superconducting qubit processors with small In bumps to accommodate a large number of connections and a small (spacing $\leq$ \SI{10}{\micro\meter}) and uniform (tilt $\leq$ \SI{100}{\micro\radian}) chip separation for scalability \cite{Kosen_2022, Norris_Michaud_Pahl_Kerschbaum_Eichler_Besse_Wallraff_2024}. However, this approach requires an elaborate fabrication process and a high-precision flip-chip bonder. Alternative flip-chip bonding techniques have been developed to mitigate this problem \cite{Satzinger_Conner_Bienfait_Chang_Chou_Cleland_Dumur_Grebel_Peairs_Povey_et_al._2019, Conner_2021}, but they do not provide a galvanic contact necessary for DC transport.

In our work, we present a simplified fabrication and assembly process for making a superconducting flip-chip connection between a superconducting Al wiring layer, which is a commonly used material for superconducting quantum circuits \cite{Rosenberg_2017, Foxen_2017, Kosen_2022, Norris_Michaud_Pahl_Kerschbaum_Eichler_Besse_Wallraff_2024}, and a superconducting niobium nitride (NbN) or niobium (Nb) layer as the UBM layer. At temperatures in the range of a few tens of millikelvin to \SI{1.2}{\kelvin}, i.e., the superconducting transition of Al, our flip-chip assembly using passivated UBM layers can carry a supercurrent of tens of \unit{\milli\ampere}, which is limited by the geometry of the patterned thin films and the current density of the utilized materials. Our approach targets applications that require only a few superconducting connections, allowing for a larger indium bump size. We demonstrate that superconducting flip-chip devices can be assembled using commercially available \SI{300}{\micro\meter} diameter indium microspheres \cite{Caplinq} as the bond connection, which, when pressed to a flat cylinder of about \SI{500}{\micro\meter} diameter, can carry a supercurrent of more than \SI{1}{\ampere}. The microspheres are placed manually on a bottom chip before pressing the bottom and a top chip together using a simple transfer stage~\cite{Zhao_2020} provided that the top chip is made of a transparent substrate (e.g., sapphire used in this work) or contains through-vias for alignment.

\begin{figure}[t!bhp]
    \centering
    \includegraphics[width=\columnwidth]{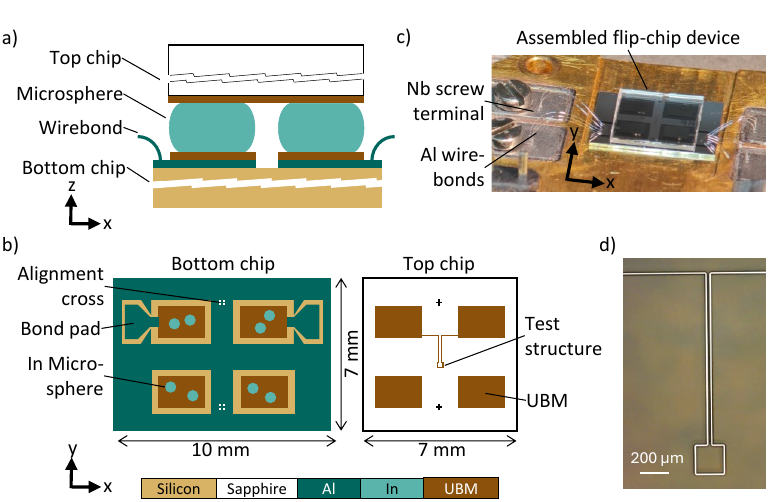}
    \caption{Schematic representation of a flip-chip device, showing (a) the side view after assembly and (b) the top view before assembly. (c) Photograph of an assembled flip-chip device wire bonded to a copper sample holder. (d) Optical image of a fabricated test structure on a sapphire substrate.}
    \label{fig:1}
\end{figure}

Our flip-chip assembly approach uses indium microspheres, eliminating the need to grow and pattern indium microbumps. To connect to a superconducting Al thin film, we demonstrate flip-chip assemblies with the commonly used NbN \cite{Gordon_1999, Kosen_2022}. Importantly, we also show that Nb can be used as a UBM layer, provided that it is capped with a thin Au layer to prevent oxidization. We choose Nb specifically because it exhibits a higher critical current density, which enables high current devices such as on-chip magnetic traps \cite{martiIEEE2022,martiPRA2023} or on-chip solenoids \cite{Graninger_Longo_Rennie_Shiskowski_Mercurio_Lilly_Smith_2019}, and its lower intrinsic microwave losses are promising for high-$Q$ superconducting resonators \cite{Verjauw_2021, Virginia_2022}.  
Au is chosen as the passivation layer because it mechanically strengthens the flip-chip bond \cite{Das_2018}, and if an Au-In intermetallic compound is formed, it would also be superconducting at millikelvin temperatures \cite{Hamilton_Raub_Matthias_Corenzwit_Hull_1965a}.


An overview of the flip-chip device is shown in \fref{fig:1}. It consists of a bottom chip made from silicon and a top chip made from sapphire that allows for in situ alignment during assembly due to its transparency. The two chips are galvanically connected through indium microspheres of \SI{300}{\micro\meter} diameter, which, after bonding, give a separation of below \SI{50}{\micro\meter}, see \fref{fig:1}(a). The bottom chip contains an Al wiring layer (\fref{fig:1}(b)), which can be used, for example, to pattern a superconducting circuit. This chip contains large areas (\SI{1.5}{\milli\meter}$\times$\SI{2}{\milli\meter}) for wire bonding to make external electrical connections to the chip. To make flip-chip interconnects, we pattern and deposit UBM pads on top of the Al layer, upon which we place indium microspheres that connect galvanically to the top chip. The top chip (\fref{fig:1}(b)) consists of a superconducting layer, which can, for example, be shaped as a loop (see \fref{fig:1}(d)) or other geometry. A completed flip-chip device is shown in \fref{fig:1}(c). The fabrication process for thin films of Al (\SI{150}{\nano\meter}), Nb (\SI{50}{\nano\meter}), NbN (\SI{50}{\nano\meter}), and Au (\SI{5}{\nano\meter}) is described in detail in the \aref{app:fab}.  

\begin{figure}[h!tbp]
    \centering
    \includegraphics[width=\columnwidth]{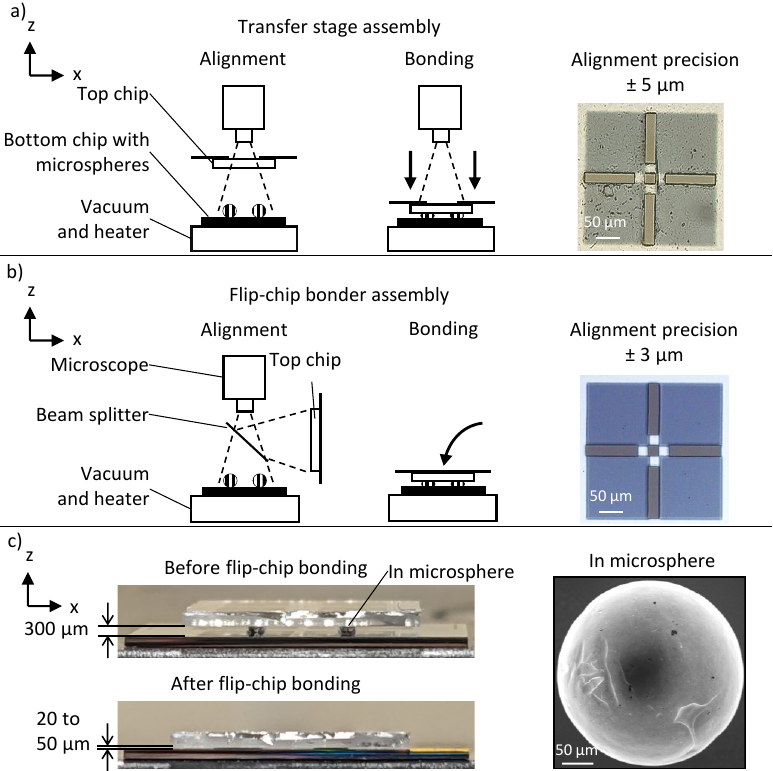}
    \caption{Schematic of the flip-chip assembly method and picture of the alignment crosses after bonding for (a) the transfer stage and (b) the flip-chip bonder. (c) Flip-chip device before and after bonding. Note that the colors on the bottom panel are light reflections of the surroundings. The close-up is an SEM image of an indium microsphere used for flip-chip bonding.}
    \label{fig:3}
\end{figure}


We assembled our flip-chip devices using two different methods: an in-house-built transfer stage (\fref{fig:3}(a)) and a commercial flip-chip bonder (\fref{fig:3}(b)). The former method is commonly used for the transfer of two-dimensional materials~\cite{Zhao_2020}. Both tools create a mechanically sturdy bond with a lateral alignment accuracy of a few micrometers. Details of the assembly procedure are described in \aref{app:assembly}. 

To make a galvanic and robust connection between the chips, we use commercially available \SI{99.99}{\percent} pure indium microspheres \cite{Caplinq} (\fref{fig:3}(c) right panel). The microspheres have a diameter of \SI{300 \pm 5}{\micro\meter} and a spheroid tolerance of \SI{1.5}{\percent}. Although In possesses a native oxide, it normally passivates at a few nanometers \cite{Schoeller2009}. This oxide layer is easily broken during bonding, as In is malleable and deforms significantly when pressed. We typically use two microspheres per pad, but in principle, a single microsphere per pad would suffice, which would then also allow the UBM pad size to be reduced, enabling us to realize more connections per chip. The applied pressure during the flip-chip bonding process flattens each \SI{300}{\micro\meter}-diameter microspheres into a flat cylinder with a diameter of around \SI{500}{\micro\meter}.

\begin{figure}[t!hbp]
    \centering\includegraphics[width=\columnwidth]{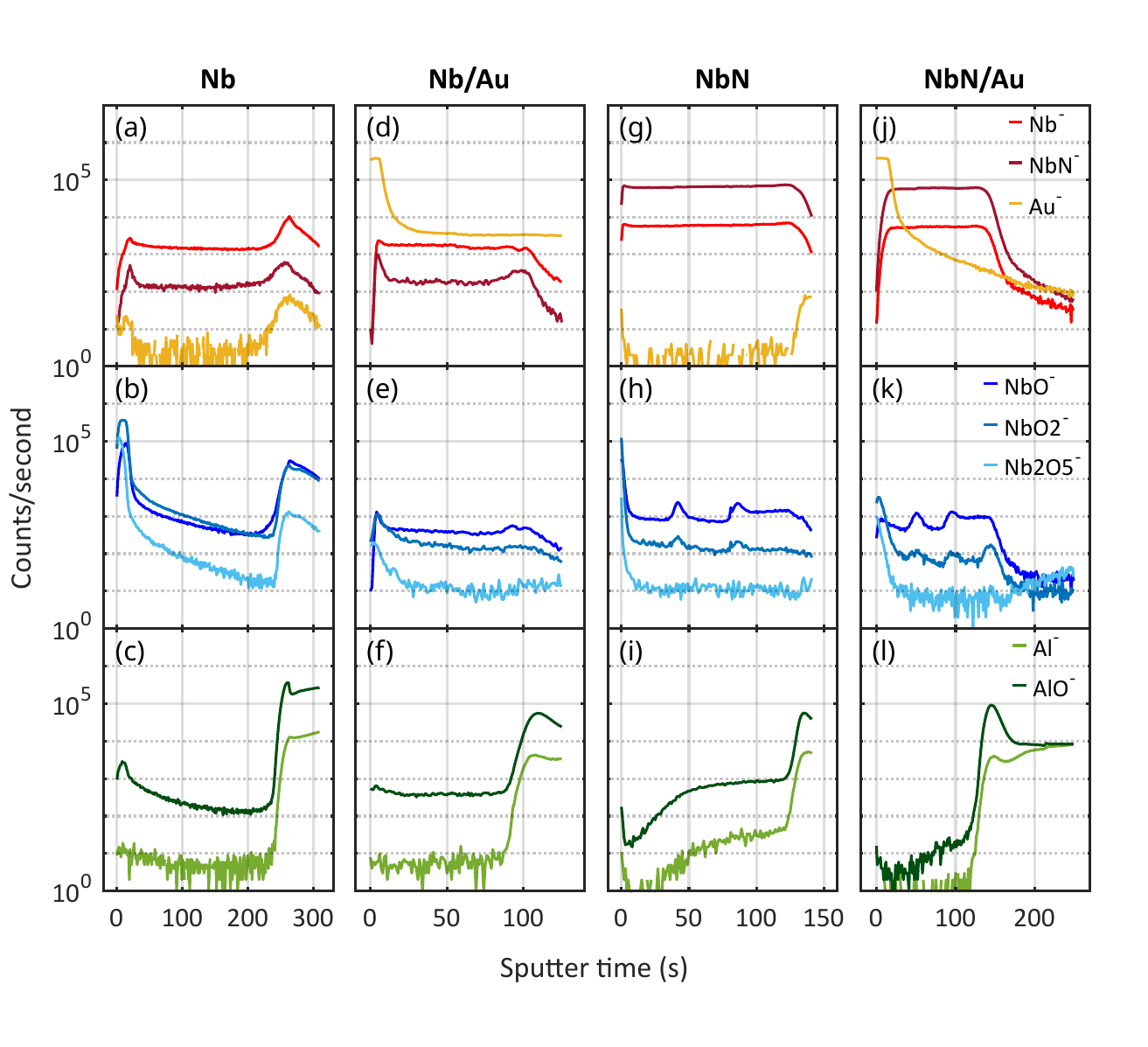}
    \caption{TOF-SIMS profiles of Nb, Nb/Au, NbN, and NbN/Au thin films acquired with \SI{1}{\second} intervals using Cs\textsuperscript{+} primary ions. The plots show the secondary ion rate as a function of sputtering time for key elements and compounds present in the samples. Top panels:  UBM metallic ions, middle panels: native Nb oxide ions, bottom panels: Al ions from Al ground plane (for Nb/Au, NbN, NbN/Au) or sapphire substrate (for Nb).}
    \label{fig:4}
\end{figure}

In general, the transfer stage achieved alignment accuracy (\SI{5}{\micro\meter}) similar to that of the flip-chip bonder (\SI{3}{\micro\meter}). However, the flip-chip bonder achieved a closer chip separation (\SI{20}{\micro\meter}) compared to the transfer stage (\SI{50}{\micro\meter}). The chip separation can be reduced in future devices using smaller indium microspheres. In addition, an average tilt of \SIrange[range-units=single]{2}{3}{\milli\radian} was obtained with both methods. The results presented in the remainder of this work were obtained using the commercial flip-chip bonder unless otherwise stated.


It is crucial to characterize the material composition and superconducting properties of the utilized UBM thin films in order to realize functioning superconducting flip-chip devices. To this end, we show measurements of the material composition of the UBM thin films using time-of-flight secondary ion mass spectrometry (TOF-SIMS) in order to analyze the potential oxidization of the non-passivated UBM thin films. \fref{fig:4} presents TOF-SIMS depth-profiles of the UBM films, in the case of Nb/Au, NbN, and NbN/Au sputtered on the Al layer of the bottom chip, and in the case of Nb sputtered directly on the sapphire substrate of the top chip. The films were analyzed a few weeks after exposure to air. The obtained SIMS data allows us to make qualitative statements about the material composition of the films (for more details, see \aref{app:sims_surface}).

We observe that non-passivated Nb and NbN films (\fref{fig:4}(b),(h)) exhibit native Nb oxides of different stoichiometry (NbO, NbO\textsubscript{2}, Nb\textsubscript{2}O\textsubscript{5}) in the surface region, rapidly decreasing in the bulk region. In contrast, Au-passivated films (\fref{fig:4}(d),(j)) show strong Au signals at the surface with minimal oxide presence (\fref{fig:4}(e),(k)), demonstrating effective passivation. The Nb and NbN layers in all samples demonstrate stable signals throughout the respective UBM film thickness (\fref{fig:4}(a),(d),(g),(j)). A sharp increase in the Al signals marks the UBM-Al interface for the Nb/Au, NbN, and NbN/Au thin films and the UBM/sapphire interface for the Nb thin film.


\begin{figure}[t!hbp]
    \centering
    \includegraphics[width=\columnwidth]{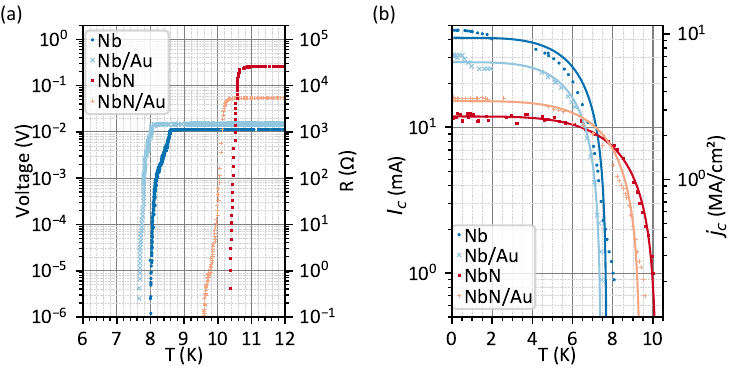}
    \caption{Measurements of superconducting thin films. (a) Voltage and resistance as a function of temperature at \SI{10}{\micro\ampere} bias current. Below the $T_c$ of the respective material the measured voltage drops to the noise floor, which has been subtracted from the shown data. (b) Critical current and critical current density as a function of temperature. The solid line is a fit using \eref{eq.1}.
    }
    \label{fig:5}
\end{figure}

To then characterize the superconducting properties of the UBM thin films, we patterned a test structure to determine the transition temperature $T_c$ and the critical current $I_c$ of the patterned passivated and non-passivated UBM thin films. The test structure is a long wire of \SI{10}{\micro\meter} width and \SI{4740}{\micro\meter} length with a loop of \SI{200}{\micro\meter} diameter in its middle (see \fref{fig:1}(d)).  We performed electrical measurements inside a dilution refrigerator, allowing us to characterize samples down to millikelvin temperature. The details of the measurement setup and procedure are explained in the \aref{app:meassetup}. 

\fref{fig:5}(a) shows the measured voltage and calculated resistance of the test structure as a function of temperature, measured with a bias current of \SI{10}{\micro\ampere}. The thin films exhibit a normal-state resistance that is in similar to values reported in literature (see \aref{app:resistance_data}). 
A sharp transition in resistance is observed at the respective $T_c$ (taken as the temperature at which the resistance is at least tenfold the residual resistance), indicating the onset of superconductivity. The respective $T_c$ for the Nb and NbN thin film is \SI{8.0}{\kelvin} and \SI{10.4}{\kelvin}, respectively, similar to the values reported in Refs.~\cite{il2008nbn,Kim_2009, Niepce_2020}. However, the $T_c$ of our Nb film is lower than the bulk value of \SI{9.3}{\kelvin} that was also observed in unpatterned thin Nb films~\cite{Joshi_2023}. We believe that this could be due to disorder in our film \cite{Freitas_Gonzalez_Nascimento_Takeuchi_Passamani_2017}, the presence of native oxides \cite{David_2014} (see \fref{fig:4}(b)), or the geometric constrictions \cite{Kim_2009} in our test structure. The Au-passivated thin films show a $T_c$ for Nb/Au and NbN/Au of \SI{7.7}{\kelvin} and \SI{9.9}{\kelvin}, respectively, slightly lower than their non-passivated counterparts. 
We attribute this small reduction in $T_c$ to the proximity effect \cite{Deutscher_1969} induced by the Au capping layer on Nb \cite{deory2024lowlosshybridnbau} or NbN \cite{il2008nbn}. 

\fref{fig:5}(b) shows the dependence of the thin film's critical current on the sample temperature. We obtain a similar $T_c$ as in the resistance measurements (\fref{fig:5}(a)). At temperatures approaching $T_c$, $I_c$ drops drastically, as expected. We fit the data assuming the simple two-fluid model \cite{Tinkham_2004a} as ($t=T/T_c$):

\begin{equation}\label{eq.1}
    I_c = I_{c0} \left( 1-t^4 \right).
\end{equation}

\begin{figure}[t!bhp]
    \centering\includegraphics[width=\columnwidth]{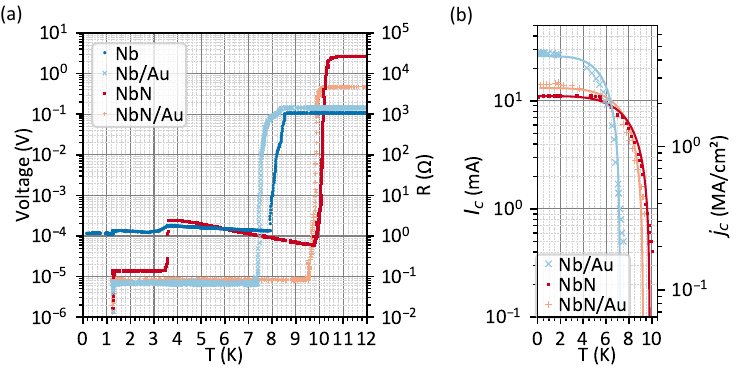}
    \caption{Measurements of flip-chip devices. (a) Voltage and resistance as a function of temperature at \SI{100}{\micro\ampere} bias current. Note that the voltage noise floor has been subtracted from the shown data. (b) Critical current and critical current density as a function of temperature.}
    \label{fig:6}
\end{figure}

This fit agrees well with the data, as seen in \fref{fig:5}(b). At a temperature of $\sim$\SI{100}{\milli\kelvin}, we determine $j_c$ of the thin films to be \SI{9.26}{\mega\ampere\per\centi\meter\squared} and \SI{2.5}{\mega\ampere\per\centi\meter\squared} for Nb and NbN, respectively, similar to other works \cite{Kim_2009,Il’in_Siegel_Semenov_Engel_Hubers_2005} (see \aref{app:resistance_data}).
The $j_c$ of Nb/Au (\SI{6.3}{\mega\ampere\per\centi\meter\squared}) and NbN/Au (\SI{3.2}{\mega\ampere\per\centi\meter\squared}) are close to their non-passivated counterparts. Note that we calculate the critical current density $j_c$ of the thin films considering the relevant geometric feature size of the superconductor, and we assume bulk transport for Nb and NbN and screening transport for Al and In. For simplicity, we neglect the influence of the proximitization from the metal layer on the current distribution in the UBM layer. Further, we overlook the possibility of sheath transport that could occur in Nb thin films.


Having characterized the UBM thin films, we then assembled flip-chip devices as described earlier. \fref{fig:6}(a) shows the measured voltage and calculated resistance of the flip-chip devices as a function of temperature. The flip-chip devices that we patterned with NbN, NbN/Au, or Nb/Au as the UBM layer were fully superconducting at millikelvin temperatures and are thus useful devices for making superconducting flip-chip-based interconnects. In contrast, the non-passivated Nb-based flip-chip device showed a finite resistance of some Ohm at millikelvin temperatures. Thus, in the following, we will focus on discussing the functioning NbN, NbN/Au, and Nb/Au-based flip-chip devices.

In \fref{fig:6}(a), we observe transitions at the $T_c$ of the respective superconducting materials. At \SI{1.25}{\kelvin}, all three flip-chip devices show a sharp increase in resistance, which marks the $T_c$ of Al. We estimate a normal-state resistivity of Al to be between \SI{0.68}{\micro\ohm\cdot\centi\meter} and \SI{1.6}{\micro\ohm\cdot\centi\meter}, which is comparable to that reported in Ref.~\cite{Sun_2017}. The resistance of flip-chip devices that use Au-passivated UBM layers remains constant between the $T_c$ of Al and another sharp transition in resistance at \SI{7.4}{\kelvin} for the Nb/Au and \SI{10.0}{\kelvin} for the NbN/Au flip-chip device, respectively. These temperatures mark the respective $T_c$ of the utilized UBM thin films and are similar to those determined for thin films only. Above their respective $T_c$, all flip-chip devices exhibit resistance values corresponding to their normal-state resistances, as observed for their respective thin films (see \aref{app:resistance_data}).

The non-passivated NbN flip-chip device shows an additional sharp increase in resistance at \SI{3.5}{\kelvin}, which corresponds to the $T_c$ of In. Given the geometry of our In microspheres, we expect their normal-state resistance to be some \SI{}{\nano\ohm}, which is well below the observed resistance change. Instead, we attribute the finite resistance to the thin native oxide layer that forms on the surface of the non-passivated NbN film due to exposure to air. \fref{fig:4}(h) shows the presence of the NbO$_x$ oxides NbO, NbO$_2$, and Nb$_2$O$_5$ in the non-passivated NbN film. In contrast, oxide growth is effectively prevented in the Au-passivated NbN thin film (see \fref{fig:4}(k)). We assume that the NbO$_x$ oxide layer acts as an insulator and estimate its resistivity to be of the order of \SI{d6}{\ohm\cdot\centi\meter} (see \aref{app:nbnnon}), 
which is reasonable for NbO$_2$ oxide\cite{Janninck_Whitmore_1966}. NbO$_2$ was also identified in Ref.~\cite{Krause_2016} as a relevant oxide of superconducting NbN thin films. We assume that below the $T_c$ of In, the NbN/NbO$_x$/In interface forms a superconductor-insulator-superconductor (SIS) connection and, thus, can transport a superconducting current below the critical current of the SIS connection. Using the Ambegaonkar-Baratoff equation, we estimate this critical current to be around \SI{2}{\milli\ampere} (see \aref{app:nbnnon}).
Close to this value, we also observe a finite resistance step in a current-dependent measurement of the NbN flip-chip device (see \aref{app:nbnnon}).
Above the $T_c$ of In, the interface should behave as a metal-insulator-superconductor (NIS) connection. We then observe that the voltage decreases with an increase in temperature, which is similar to other NIS connections~\cite{Golubev_Kuzmin_2001, Pekola_2006, Karimi_Chang_Pekola_2022}. However, to support our interpretation, additional measurements would be required to clearly indentify the non-linear behavior of the SIS or NIS junction via determining the current- and voltage-bias characteristics at different temperatures. This we leave to future work. 

\fref{fig:6}(b) shows the temperature dependence of the critical current of the functioning flip-chip devices. The critical current behavior of the flip-chip devices is governed by the smallest superconducting structure on the flip-chip device, which in our case was the test structure on the top chip's UBM layer (see \fref{fig:1}(d)). The data is well described by \eref{eq.1}, and the observed $T_c$ values agree well with those determined from the data of \fref{fig:6}(a). The flip-chip devices exhibit slightly lower $T_c$ than their thin-film counterparts, which we attribute to the proximity effect due to the Al layer \cite{Zehnder_1999} below the respective UBM layer on each bottom chip. We infer a maximum $I_c$ ($j_c$) at $\sim$\SI{100}{\milli\kelvin} of \SI{28}{\milli\ampere} (\SI{5.6}{\mega\ampere\per\centi\meter\squared}) and \SI{15}{\milli\ampere} (\SI{3.0}{\mega\ampere\per\centi\meter\squared}) for the Nb/Au and NbN/Au flip-chip devices, respectively, similar to their thin-film counterparts (see \aref{app:resistance_data}).

We also determined the critical current of the indium microsphere-based interconnects (for details, see \aref{app:indium}). 
To this end, we assembled a simplified flip-chip device using the transfer stage, with unpatterned chips of Nb/Au film chosen for its high critical current and absence of oxidation. Two In microspheres were placed on each bottom chip, while a third chip was pressed on top to establish a galvanic connection between the two bottom chips. We used 24 aluminum wire bonds per bond pad to allow the transport of large supercurrents. With this flip-chip device, we could run a supercurrent of up to \SI{2}{\ampere} (see \aref{app:indium}), 
limited by the maximum output of the current source.


To conclude, we have presented a simple approach for superconducting flip-chip devices using \SI{300}{\micro\meter} diameter indium microspheres for bonding on Au-passivated Nb- or NbN-based UBM layers. Our assembly method achieves chip separations of \SIrange[range-units=single]{20}{50}{\micro\meter} with a transversal alignment accuracy of better than \SI{5}{\micro\meter} and a tilt of \SIrange[range-units=single]{2}{3}{\milli\radian} using either a transfer stage or a conventional flip-chip bonder. Smaller chip separations may be achieved using smaller indium microspheres, which would require a smaller pad size. This would also allow for patterning a larger number of pads, thus resulting in more superconducting interconnects per chip.

Our flip-chip assembly is suitable for high-current applications. The indium interconnects, with a pressed diameter of about \SI{500}{\micro\meter}, can carry currents exceeding \SI{1}{\ampere} at millikelvin temperatures. With patterned superconducting thin films, we have demonstrated that our flip-chip devices could run tens of \unit{\milli\ampere} at temperatures below the superconducting transition of Al, which was used as a wiring layer in the bottom chip. 
We found that although native oxides were present in both non-passivated Nb and NbN flip-chip devices, only the latter were superconducting at millikelvin temperatures. An Au passivation layer enabled functioning Nb flip-chip devices, opening up the possibility for high-current applications since the critical current density of Nb is larger than that of NbN or Al.

The presented flip-chip devices are suitable for various superconducting devices, such as chip-based high-current magnetic traps \cite{martiIEEE2022, martiPRA2023} or solenoids \cite{Graninger_Longo_Rennie_Shiskowski_Mercurio_Lilly_Smith_2019}, efficient flux readout in SQUID-based sensors \cite{martiSUST2020,martiPRA2023,Schmidt2024}, or transport of supercurrent to flux-tunable superconducting couplers \cite{Chen_Neill_Roushan_Leung_Fang_Barends_Kelly_Campbell_Chen_Chiaro_etal._2014, Niu_Gao_He_Wang_Wang_Lin_2024}. Thus, our flip-chip devices can find applications in superconducting-based sensors \cite{DeNigris_Chervenak_Bandler_Chang_Costen_Eckart_Ha_Kilbourne_Smith_2018, Lucas_Biesecker_Doriese_Duff_Hilton_Ullom_Vissers_Schmidt_2022} or superconducting quantum technologies \cite{Foxen_2017, Rosenberg_2017, Conner_2021, Kosen_2022, Norris_Michaud_Pahl_Kerschbaum_Eichler_Besse_Wallraff_2024, Schmidt2024}.

\begin{acknowledgments}
		The authors thank Marcus Rommel and Henrik Frederiksen for their support in microfabrication, Hanlin Fang and Thilo Bauch for technical support, and Per Malmberg for performing TOF-SIMS measurements. This work was supported in part by the Horizon Europe 2021-2027 framework program of the European Union under Grant Agreement No.~101080143 (SuperMeQ), the European Research Council under Grant No.~101087847 (ERC Consolidator SuperQLev), the Knut and Alice Wallenberg (KAW) Foundation through a Wallenberg Academy Fellowship (WW), and the Wallenberg Center for Quantum Technology (WACQT, AP). The devices were fabricated at Chalmers Myfab Nanofabrication Laboratory and analyzed in part at Chalmers Materials Analysis Laboratory (CMAL).
\end{acknowledgments}

Data underlying the results presented in this paper are available in the open-access Zenodo database: \href{https://doi.org/10.5281/zenodo.13377107}{10.5281/zenodo.13377107}.



\addcontentsline{toc}{section}{Supplementary Material}
\section*{Supplementary Material}

\beginsupplement

\section{Fabrication}\label{app:fab}

\begin{figure}[t!hbp]
    \centering
    \includegraphics[width=\columnwidth]{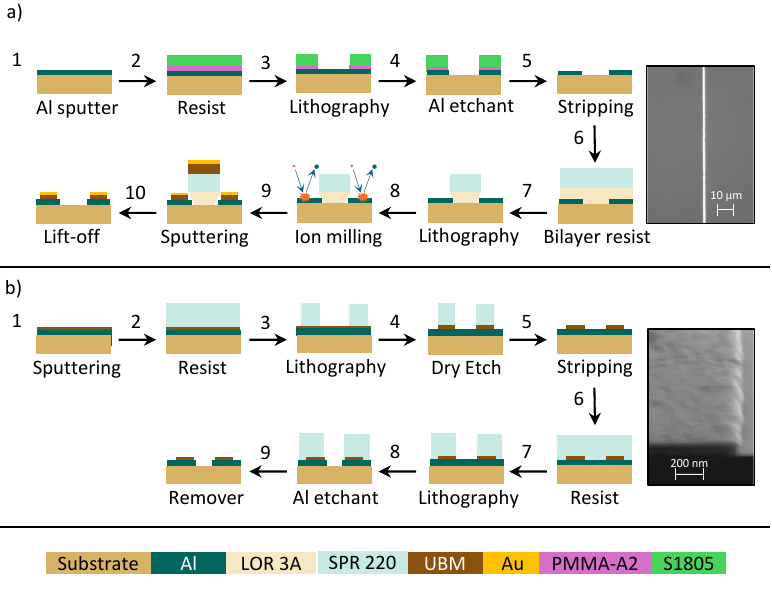}
    \caption{(a) Schematic of the liftoff process. Right: an optical image of a \SI{2}{\micro\meter} wide test structure. (b) Schematic of the etching process. Right: SEM image of a \SI{100}{\nano\meter} thick UBM film.}
    \label{sfig:1}
\end{figure}

In the following, we describe the fabrication of the layers on the bottom chip since it contains the UBM layer and the additional Al wiring layer. We present two approaches for patterning the UBM layer: a liftoff process and an etching process. The top chip consists of only the UBM layer patterned with pads and the test structure, and thus, the fabrication procedure is the same as that for the bottom chip without the process steps of the Al layer. Before any film is deposited, the wafers are cleaned in an ultrasonic bath with acetone, followed by isopropyl alcohol (IPA).

The liftoff process (\fref{sfig:1}(a)) begins with the fabrication of an Al wiring layer. First, \SI{150}{\nano\meter} of Al is sputtered on a 2" Si wafer using a DC magnetron sputtering tool (step 1). The deposition is followed by the spin coating of resists PMMA-A2 (to protect Al against developer MF-319) and S1805  (step 2), exposing the pattern with optical lithography and developing the resist in MF-319 (first part of step 3). The PMMA-A2 in the developed area is then ashed in O$_2$ plasma at \SI{40}{\watt} for \SI{90}{\second}  (second part of step 3). The Al layer is then etched in Transene Etchant A at \SI{40}{\degreeCelsius} for \SI{90}{\second} and rinsed with water  (step 4). Finally, the resist is stripped with Remover 1165  (step 5), and the wafer is cleaned as mentioned above. The UBM layer is fabricated by spin-coating the wafer with a bilayer resist (LOR 3A (\SI{0.4}{\micro\meter}) \& SPR 220 (\SI{3}{\micro\meter})) to create an undercut and avoid sidewall deposition for easier liftoff  (step 6). The pattern is then exposed using optical lithography and developed in MF-24A. The wafer is flood-exposed using a UV lamp to prevent bubble formation during sputtering due to trapped N$_2$, followed by descumming in O$_2$ plasma as before  (step 7). To deposit the UBM layer, we use a magnetron sputtering system with an Nb DC source at a base pressure of $\sim$\SI{d-8}{\milli\bar}. Before sputtering, we ion-milled the surface  (step 8) with an Ar flow of \SI{2}{sccm} and a beam current of \SI{10}{\milli\ampere} for \SI{50}{\second} to remove native Al oxide for galvanic contact between the wiring layer and the UBM pads. The DC power source is set at \SI{200}{\watt} during the deposition of Nb and NbN  (first part of step 9) and sputtered in steps of \SI{17}{\nano\meter} to avoid excessive heating and therefore bubbling of the resist during deposition. Additionally, for depositing NbN, the N$_2$ flow is set at \SI{8}{sccm} and the Ar flow to \SI{60}{sccm} for a stoichiometry of approximately 1:1. Surface profilometry shows that the thickness uniformity of the Nb or NbN film is precise to within \SI{10}{percent} of the desired thickness. A thin layer of \SI{5}{\nano\meter} Au is sputtered in the same vacuum for devices with passivation (second part of step 9). Finally, the wafer is soaked overnight in Remover 1165 for liftoff of the UBM layer (along with the Au layer, if present)  (step 10) and then cleaned with acetone and IPA.

In the etching process (\fref{sfig:1}(b)), the Al wiring layer and the UBM layer (Nb or NbN) are sputtered in the same vacuum  (step 1), one after the other, on a pre-cleaned bare wafer. The UBM resist mask  (step 2) is then patterned on top of the thin film using SPR220 and optical lithography. After the resist is developed  (step 3), the wafer is placed in a reactive ion etching chamber. The UBM layer is selectively etched with NF$_3$ gas  (step 4), which does not attack the Al layer. The resist is dissolved in Remover 1165  (step 5), and a second resist mask of SPR220  (step 6) is patterned to define the wiring layer. After development and ashing  (step 7), the wafer is submerged in a Transene etchant bath to etch the Al  (step 8). Finally, the resist mask is dissolved in Remover 1165  (step 9), and the wafer is cleaned as stated above.

Both processes are suitable alternatives for fabricating the desired material stacks. In some devices, the adhesion of Au to Nb was inadequate. Future processes could mitigate this issue by incorporating an additional \SI{5}{\nano\meter} layer of Ti as an adhesion promoter \cite{Das_2018}. With liftoff, we have reliably patterned long and narrow features with widths as small as \SI{2}{\micro\meter} using optical lithography (\fref{sfig:1}(a)) and can, in principle, pattern sub-\unit{\micro\meter} features using e-beam lithography. Patterning such features via etching requires a high degree of control over selectivity, anisotropy, and etch rate. Therefore, the liftoff process is preferred to pattern narrow and long coil features \cite{martiIEEE2022,martiPRA2023}, coplanar waveguides \cite{Verjauw_2021}, or microbridges \cite{Haberkorn_Lee_Lee_Yun_Verón_Lagger_Sirena_Kim_2024, Du_Xu_Zhang_Li_Wei_Wang_Hou_Chen_Liu_Liu_et_al._2024}. However, liftoff is viable only up to a certain film thickness, beyond which the resist may harden during deposition and become insoluble, or the resist may not be thick enough. Thus, an etching-based process is preferred for layers thicker than \SI{100}{\nano\meter} (\fref{sfig:1}(b)). Certain deposition tools lack ion milling capability but can deposit multiple materials within the same vacuum environment. In that case, the etching process can be used to pattern each film successively with suitable selective etching processes.

\section{Flip-chip assembly}\label{app:assembly}

In our first approach to flip-chip bonding (\ffref{fig:3}(a)), we use an $xyz$ micro-manipulator stage. The bottom chip is held via vacuum, and the indium microspheres are placed on the UBM pads with tweezers.  The top chip is attached to the glass slide with polydimethylsiloxane (PDMS) tape and is positioned above the bottom chip. The two chips are aligned with respect to each other via alignment markers, which can be seen through the transparent sapphire substrate of the top chip in a microscope. Alternatively, non-transparent substrates could also be used if they have through-vias for alignment. The bottom chip can be heated to \SI{110}{\degreeCelsius} if reflow of indium is desired; otherwise, it can be kept at room temperature. The reflow temperature must remain well below \SI{156}{\degreeCelsius} to avoid melting of indium, and also low enough to prevent oxide growth that can hinder galvanic contact with the UBM layers~\cite{Schoeller2009}. Finally, the top chip is brought down and pressed against the indium microspheres using the manipulator stage. Since the applied force is unknown, the pressing is stopped if any relative lateral motion between the two chips is observed. Additionally, the $x$-$y$ alignment is continuously monitored and corrected during the bonding procedure.

We have also used a commercial flip-chip bonder (Fineplacer 96 Lambda from Finetech GmbH), \ffref{fig:3}(b), to bond the chips and compare performance with the manual transfer stage. The top chip is held with a vacuum on a pick-up tool that stands vertically next to the bottom chip. A beam splitter in the optical path enables both chips to be seen simultaneously. Once alignment is achieved, the microscope is retracted, and a precalibrated lever presses the top chip onto the bottom chip with a calibrated force (bonding pressure) between \SIrange[range-units=single]{50}{65}{\newton} (\SIrange[range-units=single]{60}{80}{\mega\pascal}) at room temperature. The alignment cannot be monitored or corrected during bonding.

\section{Experimental setup and measurement procedure}\label{app:meassetup}

\begin{figure}[t!bhp]
\centering\includegraphics[width=\columnwidth]{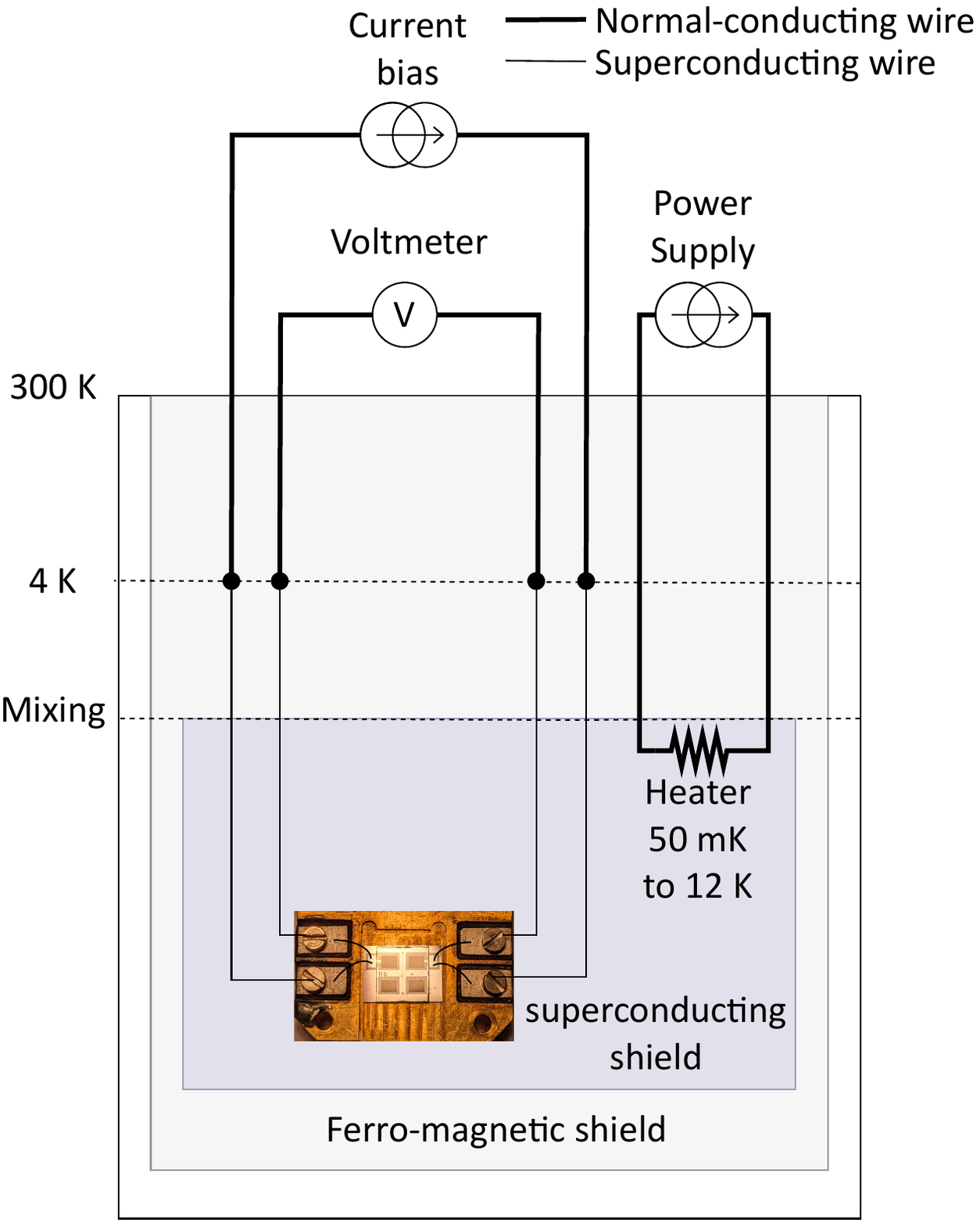}
    \caption{Schematic of the four-wire measurement setup inside the dilution refrigerator. The sample is wire-bonded to the Nb screw terminals of the copper holder. The screw terminals clamp superconducting NbTi wires that terminate at the 4\,K plate, where the normal-conducting wires make connections to the instruments. A controlled resistive heater is located on the mixing chamber plate (base plate) to vary the temperature of the sample.}
    \label{fig:setup}
\end{figure}

We performed electrical measurements inside a dilution refrigerator (see \fref{fig:setup}), allowing us to characterize samples to millikelvin temperature. The chips are glued with BF-6 glue to a copper sample holder, as shown in \ffref{fig:1}(c). The test structure bond pads are connected to Nb screw terminals on a sample holder with Al wirebonds (\SI{25}{\micro\meter} diameter).

Below the $T_c$ of Al, each Al bond wire can carry a supercurrent of up to \SI{85}{\milli\ampere}. The screw terminals are used to clamp twisted niobium-titanium (Nb-Ti) wires that are soldered to a DC breakout on the \SI{4}{\kelvin} plate of the cryostat. Thus, the entire line is superconducting below the \SI{4}{\kelvin} plate. 

Electrical characterization is performed by measuring the resistance of a four-wire connection to the device under test, which excludes the wire bonds and other connections to the sample. The bias current is provided with a Keithley 2450 source, and voltage is measured by a Rohde \& Schwarz HMC8012. We subtracted a systematic non-zero voltage offset of the voltmeter from the data presented in \ffref{fig:5}(a) and \ffref{fig:6}(a). The HMC8012 has a systematic voltage offset uncertainty of $\pm$\SI{8}{\micro\volt}, which translates to $\pm$\SI{0.8}{\ohm} ($\pm$\SI{0.08}{\ohm}) for a bias current of $10\,\mu$A ($100\,\mu$A).

We controlled the temperature of the samples by varying the power to an inbuilt heater on the base plate of the dilution refrigerator in the range of \SI{50}{\micro\watt} to \SI{1}{\milli\watt} while circulating the He-3/He-4 mixture, and further up to \SI{0.8}{\watt} with only pre-cooling turned on.

For the measurements to determine the critical current at various temperatures, we infer the $I_c$ at each temperature from individual measurements, in which we increase the current in steps of \SI{0.1}{\micro\ampere} until the measured resistance increases sharply as a result of the transition from the superconducting state to the normal conducting state. An increase in the temperature of the cryostat accompanies this increase in resistance. Before taking the next measurement, we set the power to a built-in heater on the base plate of the dilution refrigerator and wait a sufficient amount of time for the cryostat to thermalize to the desired temperature.

The contribution of Al bond pads to the resistance and the critical current values is negligible due to their large cross-sectional area (\SI{1}{\milli\meter} $\times$ \SI{150}{\nano\meter}) and surface sheath area (determined by $\lambda_L$ = \SI{54}{\nano\meter} \cite{lopez_2023}) compared to the test structure patterned in the UBM layer.

\section{TOF-SIMS}\label{app:sims_surface}

Time-of-flight secondary ion mass spectrometry (TOF-SIMS) measurements were performed using Cs$^+$ primary ions that sputter the material with an energy of \SI{1}{\kilo\electronvolt} over a \SI{300}{\micro\meter} $\times$ \SI{300}{\micro\meter} area for Nb thin film and \SI{250}{\micro\meter} $\times$ \SI{250}{\micro\meter} area for the remaining thin films. The created secondary ions are sputtered at varying rates depending on the secondary ion's mass and ionization potential. The SIMS analyzer measures the secondary ions, recording counts as a function of sputtering time.

When examining the NbN UBM oxide data in \ffref{fig:4}(h,k), we observe peaks in the Nb oxide signals. It is important that the height of these peaks is small compared to the background Nb oxide signal and much smaller than the initial Nb oxide signal close to the surface. We hypothesize that these peaks could result from changes in the interfacial layer that cause localized oxygen enrichment at specific depths. Residual oxygen from sample exposure to air and trace amounts in the vacuum system during sputtering may also contribute to the observation of such peaks. 

Furthermore, we observe a decrease in the Nb oxide signal in \ffref{fig:4}(b). We thereby distinguish two behaviors: a rapid decrease of the signal at the surface and a slow decrease within the film. The former rapid decrease we attribute to surface oxidation of the Nb film, which we explain in the main text. The slow decrease within the film is not so clear to us. This could be due to variations in film homogeneity or sample preparation artifacts, such as grains or pores, which can affect oxygen distribution and lead to different signal behaviors.

\section{Collected data: resistance, transition temperature, critical current density}\label{app:resistance_data}

\subsection{UBM materials and flip-chip devices}

We summarize the measured resistance $R$, the calculated resistivity $\rho$ (both determined at \SI{12}{\kelvin} and \SI{300}{\kelvin}), and the calculated residual resistance ratio $RRR=R_{\SI{300}{\kelvin}}/R_{\SI{12}{\kelvin}}$ for the UBM thin films in \tref{thin-film_results:resistance} and for the flip-chip devices in \tref{flip-chip_results:resistance}.

\begin{table*}[t!hbp]    
    \begin{tabular}{@{}lccccccc@{}}
        \toprule
        \parbox[t]{3cm}{\centering Thin film\\ material [thickness]} &
        \multicolumn{1}{c}{\parbox[t]{1cm}{\centering $R_{\SI{12}{\kelvin}}$\\ (\unit{\kilo\ohm})}} &
        \multicolumn{1}{c}{\parbox[t]{1cm}{\centering $R_{\SI{300}{\kelvin}}$\\ (\unit{\kilo\ohm})}} &
        \multicolumn{1}{c}{\parbox[t]{1cm}{\centering $\rho_{\SI{12}{\kelvin}}$\\ (\unit[inter-unit-product=\cdot]{\micro\ohm\cdot\centi\meter})}} &
        \multicolumn{2}{c}{\parbox[t]{2cm}{\centering $\rho_{\SI{300}{\kelvin}}$\\ (\unit[inter-unit-product=\cdot]{\micro\ohm\cdot\centi\meter})}} &
        \multicolumn{2}{c}{\parbox[t]{2cm}{\centering $RRR$}} \\
        \cmidrule(lr){5-6} \cmidrule(lr){7-8}
        & & & & Exp. & Lit. & Exp. & Lit. \\
        
        \midrule
        Nb[\SI{50}{\nano\meter}] & 1.13 & 2.89 & 11.9 & 30.5 & 19.4 \cite{Verjauw_2021} & 2.6 & 6.5 \cite{Verjauw_2021} \\
        
        Nb[\SI{50}{\nano\meter}] / Au[\SI{5}{\nano\meter}] & 1.47 & 2.54 & 17.1 & 29.5 & - & 1.73 & - \\
        
        NbN[\SI{50}{\nano\meter}] & 26.1 & 21.9 & 278 & 233 & $\sim$250 \cite{Glowacka2014-vq} & 0.84 & $\sim$0.8 \cite{Glowacka2014-vq}\\

        NbN[\SI{50}{\nano\meter}] / Au[\SI{5}{\nano\meter}] & 5.41 & 6.50 & 62.9 & 75.6 & - & 1.2 & -\\
        \botrule
    \end{tabular}
    \caption{Electrical properties of the sputtered UBM layers patterned independently on a sapphire substrate. The literature values are taken from works that use similar deposition techniques and film thicknesses.}
    \label{thin-film_results:resistance}
\end{table*}

\begin{table*}[t!hbp]
\begin{tabular}{@{}lccccc@{}}
\toprule
\parbox[t]{3cm}{\centering UBM \\ material [thickness]} &
\parbox[t]{1cm}{\centering $R_{\SI{12}{\kelvin}}$\\ (\unit{\kilo\ohm})} &
\parbox[t]{1cm}{\centering $R_{\SI{300}{\kelvin}}$\\ (\unit{\kilo\ohm})} &
\parbox[t]{2cm}{\centering $\rho_{\SI{12}{\kelvin}}$\\ (\unit[inter-unit-product=\cdot]{\micro\ohm\cdot\centi\meter})} &
\parbox[t]{2cm}{\centering $\rho_{\SI{300}{\kelvin}}$\\ (\unit[inter-unit-product=\cdot]{\micro\ohm\cdot\centi\meter})} &
\parbox[t]{1cm}{\centering $RRR$} \\

\midrule
Nb[\SI{50}{\nano\meter}] / Au[\SI{5}{\nano\meter}] & 1.42 & 2.66 & 16.52 & 30.86 & 1.87 \\
NbN[\SI{50}{\nano\meter}] & 27.2 & 21.7 & 286.7 & 228.9 & 0.81 \\
NbN[\SI{50}{\nano\meter}] / Au[\SI{5}{\nano\meter}] & 4.69 & 5.80 & 54.5 & 67.3 & 1.23 \\
\botrule
\end{tabular}
\caption{Electrical properties of the flip-chip devices with different UBM materials.}
\label{flip-chip_results:resistance}
\end{table*}

\begin{table*}[t!hbp]
    \begin{tabular}{@{}lcccccc@{}}
        \toprule
        \parbox[t]{3cm}{\centering Thin film \\ material [thickness]} &
        \multicolumn{2}{c}{\parbox[t]{2cm}{\centering $j_c$\\ (\unit[per-mode=symbol, inter-unit-product=\text{/}]{\mega\ampere\per\centi\meter\squared})}} &
        \multicolumn{2}{c}{\parbox[t]{2cm}{\centering $T_c$ \\ (\unit{\kelvin})}} \\
        \cmidrule(lr){2-3} \cmidrule(lr){4-5}
        & Exp. & Lit. & Exp. & Lit. \\
        
        \midrule
        Nb[\SI{50}{\nano\meter}] & 9.3 & 4 \cite{Kim_2009} & 8.0 & 8.1 \cite{Kim_2009} \\
        
        Nb[\SI{50}{\nano\meter}] / Au[\SI{5}{\nano\meter}] & 6.3 & - & 7.7 & 7.3 \cite{deory2024lowlosshybridnbau} \\
        
        NbN[\SI{50}{\nano\meter}] & 2.5 & 2.5 \cite{Il’in_Siegel_Semenov_Engel_Hubers_2005} & 10.4 & 10.6 \cite{Niepce_2020} \\

        NbN[\SI{50}{\nano\meter}] / Au[\SI{5}{\nano\meter}] & 3.2 & - & 9.9 & 10.2 \cite{il2008nbn} \\
        \botrule
    \end{tabular}
    \caption{Superconducting properties of the sputtered UBM layers patterned independently on a sapphire substrate. The literature values are taken from works that use similar deposition techniques and film thicknesses.}
    \label{thin-film_results}
\end{table*}

\begin{table*}[t!hbp]
\begin{tabular}{@{}lcc@{}}
\toprule
\parbox[t]{3cm}{\centering UBM \\ material [thickness]} &
\parbox[t]{2cm}{\centering $j_c$\\ (\unit[per-mode=symbol,inter-unit-product=\text{/}]{\mega\ampere\per\centi\meter\squared})} &
\parbox[t]{1cm}{\centering $T_c$\\ (\unit{\kelvin})} \\

\midrule
Nb[\SI{50}{\nano\meter}] / Au[\SI{5}{\nano\meter}] & 5.6 & 7.4 \\
NbN[\SI{50}{\nano\meter}] & 0.4$^*$ & 10.0 \\
NbN[\SI{50}{\nano\meter}] / Au[\SI{5}{\nano\meter}] & 3.0 & 9.7 \\
\botrule
\end{tabular}
\caption{Superconducting properties of the flip-chip devices with different UBM materials. $^*$The NbN flip-chip device exhibits \SI[per-mode=symbol, inter-unit-product=\text{/}]{2.2}{\mega\ampere\per\centi\meter\squared} limited by the NbN test structure, and \SI[per-mode=symbol, inter-unit-product=\text{/}]{0.4}{\mega\ampere\per\centi\meter\squared} limited by Nb oxide in the flip-chip interconnects.}
\label{tab:flip-chip_results}
\end{table*}

The $\rho_{\SI{300}{\kelvin}}$ and $RRR$ of the non-passivated thin films align with Refs.~\cite{Verjauw_2021, Glowacka2014-vq}. The Au-passivated films exhibit lower $\rho_{\SI{300}{\kelvin}}$ values compared to their non-passivated counterparts due to the parallel conductance of the Au layer.

To verify the effect of the Au film, we estimate $\rho_{\SI{300}{\kelvin}}$ of \SI{5}{\nano\meter} Au to be \SI{11}{\micro\ohm\cdot\centi\meter} \cite{SALVADORI_VAZ_FARIAS_CATTANI_2004}, and obtain $\rho_{\SI{300}{\kelvin}}$ of \SI{15}{\micro\ohm\cdot\centi\meter} for Nb and \SI{69}{\micro\ohm\cdot\centi\meter} for NbN, close to our measured values. Similarly, using $\rho_{\SI{12}{\kelvin}}$ of Au to be \SI{2.6}{\micro\ohm\cdot\centi\meter} \cite{Sambles_Elsom_Jarvis_1982}, we calculate $\rho_{\SI{12}{\kelvin}}$ of \SI{9}{\micro\ohm\cdot\centi\meter} for Nb and \SI{26}{\micro\ohm\cdot\centi\meter} for NbN.

The corresponding values of the flip-chip devices are very close to those of their thin-film counterparts, with slightly lower $R$, and therefore $\rho$, due to the parallel resistance of the Si substrate at room temperature. 

\tref{thin-film_results} and \tref{tab:flip-chip_results} summarize the determined $T_c$ and $j_c$ for the UBM thin films and the flip-chip devices, respectively, and compare our values to the values in the literature.

\subsection{Aluminum thin film}
The resistance of the Al film is obtained from the resistance step observed at the $T_c$ of Al (\ffref{fig:6}(a)) to be about \SIrange[range-units=single]{0.08}{0.2}{\ohm}. Considering the smallest cross-sectional area in the Al layer (\SI{300}{\micro\meter} $\times$ \SI{150}{\nano\meter} of length \SI{400}{\micro\meter}), we estimate that the normal-state resistivity of Al lies between \SI{0.68}{\micro\ohm\cdot\centi\meter} and \SI{1.6}{\micro\ohm\cdot\centi\meter}, which is comparable to that reported in Ref.~\cite{Joshi_2005}.

\subsection{Indium microsphere}
Bulk indium exhibits an electrical resistivity of \SI{1.2}{\nano\ohm\cdot\centi\meter} at \SI{4}{\kelvin} \cite{Hall_1968}. Approximating a pressed indium microsphere as a cylinder with \SI{500}{\micro\meter} diameter and \SI{50}{\micro\meter} height yields an estimated normal-state resistance of \SI{30}{\nano\ohm}.

\section{Non-passivated NbN flip-chip device}\label{app:nbnnon}

\subsection{Critical current of the SIS junction}

In the non-passivated NbN flip-chip device, we observe a resistance step of \SI{2.5}{\ohm} at the transition temperature of In (see \ffref{fig:6}(a)). We assume that this resistance comes from the oxide on top of NbN. Considering a contact area with a diameter of about \SI{500}{\micro\meter} (for pressed microspheres) and an oxide thickness of \SI{0.5}{\nano\meter} (estimated from the SIMS data, \ffref{fig:4}(g), assuming a sputtering rate of \SI{3.6}{\nano\meter\per\min} \cite{He_Xu_Foroughi‐Abari_Karpuzov_2012}) we obtain a resistivity of the order of \SI{d6}{\ohm\cdot\centi\meter}, which is expected for NbO$_2$ at such low temperatures \cite{Janninck_Whitmore_1966}.

The total resistance $R$ = \SI{2.5}{\ohm} is assumed to be evenly distributed at each oxide interface of the entire flip-chip device. Since the In microsphere is pressed with NbN on top and bottom, we consider two oxide interfaces in series per microsphere. Considering two microspheres per pad implies that the normal-state resistance at each interface is $R_N=R/2= \SI{1.25}{\ohm}$. 

Using the Ambegaonkar-Baratoff equation \cite{Tinkham_2004a}, we can then estimate the critical current of the SIS junction formed by the NbN/NbO$_x$/In interface as

\begin{equation}
    I_{c0} = \frac{\pi \sqrt{\Delta_{In}\Delta_{NbN}}}{2e R_N} \tanh\left(\frac{\sqrt{\Delta_{In}\Delta_{NbN}}}{2k_{\mathrm{B}}T}\right),
\end{equation}
where $k_{\mathrm{B}}$ is Boltzmann's constant, $e$ electric charge, $T$ temperature (\SI{100}{\milli\kelvin}), the superconducting energy gaps of NbN $\Delta_{NbN}$ = \SI{4.16}{\milli\electronvolt} \cite{Komenou_Yamashita_Onodera_1968} and In $\Delta_{In}$ = \SI{1.05}{\milli\electronvolt} \cite{Averill_Straus_Gregory_1972}. We obtain $I_{c0}$ = \SI{2.6}{\milli\ampere}.

\subsection{Current measurement}

\begin{figure}[t!hbp]
\centering\includegraphics[width=\columnwidth]{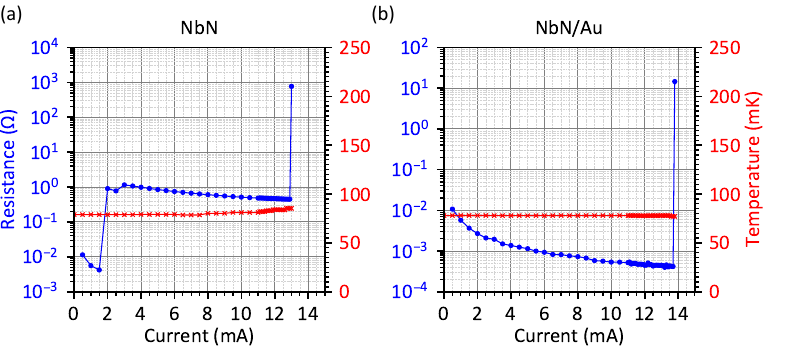}
    \caption{Current-dependent measurements of (a) non-passivated NbN and (b) Au-passivated NbN flip-chip devices. The plots show resistance (blue circles) and temperature at the mixing chamber (red crosses) as a function of current. Lines are a guide to the eye.}
    \label{fig:7}
\end{figure}

\fref{fig:7} shows resistance vs.~current measurements of the non-passivated NbN and Au-passivated NbN flip-chip devices. For both flip-chip devices, we observe a sharp increase in resistance above a current of \SI{12}{\milli\ampere}, which indicates the critical current through the test structure of the NbN thin film. For the non-passivated NbN flip-chip device (\fref{fig:7}(a)), we observe an additional resistance step of about \SI{1}{\ohm} starting at a current close to \SI{2}{\milli\ampere}. We attribute this additional resistance step to the presence of the interface oxide in the flip-chip interconnect, which forms an SIS junction with about this critical current, which is close to the estimated \SI{2.6}{\milli\ampere} from the Ambegaonkar-Baratoff equation. 

\section{Critical current of indium microsphere}\label{app:indium}

\begin{figure}[t!hbp]
    \centering\includegraphics[width=\columnwidth]{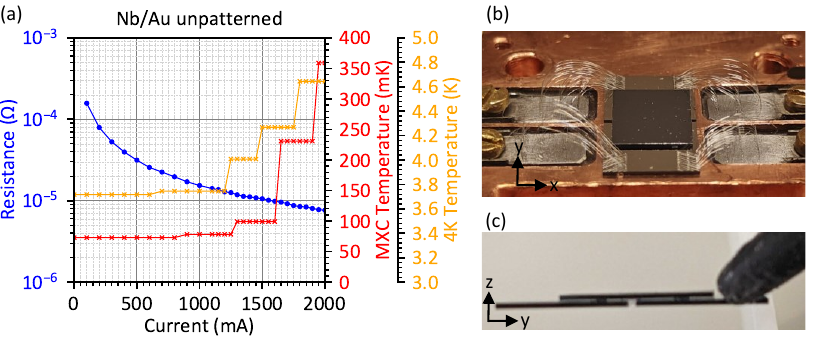}
    \caption{(a) Current measurement of an unpatterned Nb/Au flip-chip device using In microsphere interconnects. The plot shows resistance (blue curve) as well as the temperature at the base plate (red curve) and 4\,K stage (orange curve) in dependence on current. Image of (b) the assembled Nb/Au flip-chip device and (c) a close-up of its cross-section. The flip-chip device was wire bonded with 24 Al bonds per pad.}
    \label{fig:8}
\end{figure}

We assembled a simplified flip-chip device using the transfer stage, with unpatterned chips of Nb/Au film chosen for its high critical current and absence of oxidation. Two In microspheres were placed on each bottom chip, while a third chip was pressed on top to establish a galvanic connection between the two bottom chips \fref{fig:8}(c). We used 24 aluminum wire bonds per bond pad (\fref{fig:8}(b)) to allow the transport of large supercurrents.

The critical current of two Indium microspheres is measured to be above \SI{2}{\ampere} (see \fref{fig:8}(a)). In our measurement, we were limited by the maximum output of our current source, which was \SI{2}{\ampere}. Assuming a diameter ($d$) of the contact area of the pressed In microsphere of \SI{500}{\micro\meter} and the current to be flowing only within the London penetration depth of \SI{40}{\nano\meter} \cite{Anderson_1972} we estimate the $j_c$ of In to be greater than \SI [per-mode=symbol]{1.6}{\mega\ampere\per\centi\meter\squared}. Considering a critical field ($H_c$) of \SI{2.75d-2}{\tesla} \cite{Maxwell_1954} for pure bulk In, we estimate the critical current ($I_c$) using the Silsbee criterion ($I_c = \pi d H_c$ \cite{Tinkham_2004a}) to be about \SI{34}{\ampere}, which is higher than the value we can measure with the available current source. The observed heating in the cryostat for currents larger than about \SI{1.2}{\ampere} is the result of Joule heating at the 4\,K stage, where the superconducting NbTi connection terminates in a normal copper connection. During measurement, this connection heats up, which is seen as a rise in temperature at the 4\,K stage in \fref{fig:8}(a). This heat propagates down to the sample, which is seen as an increase in the base temperature at the mixing chamber.

\bibliography{sn-bibliography}

\begin{thebibliography}{60}%
\makeatletter
\providecommand \@ifxundefined [1]{%
 \@ifx{#1\undefined}
}%
\providecommand \@ifnum [1]{%
 \ifnum #1\expandafter \@firstoftwo
 \else \expandafter \@secondoftwo
 \fi
}%
\providecommand \@ifx [1]{%
 \ifx #1\expandafter \@firstoftwo
 \else \expandafter \@secondoftwo
 \fi
}%
\providecommand \natexlab [1]{#1}%
\providecommand \enquote  [1]{``#1''}%
\providecommand \bibnamefont  [1]{#1}%
\providecommand \bibfnamefont [1]{#1}%
\providecommand \citenamefont [1]{#1}%
\providecommand \href@noop [0]{\@secondoftwo}%
\providecommand \href [0]{\begingroup \@sanitize@url \@href}%
\providecommand \@href[1]{\@@startlink{#1}\@@href}%
\providecommand \@@href[1]{\endgroup#1\@@endlink}%
\providecommand \@sanitize@url [0]{\catcode `\\12\catcode `\$12\catcode
  `\&12\catcode `\#12\catcode `\^12\catcode `\_12\catcode `\%12\relax}%
\providecommand \@@startlink[1]{}%
\providecommand \@@endlink[0]{}%
\providecommand \url  [0]{\begingroup\@sanitize@url \@url }%
\providecommand \@url [1]{\endgroup\@href {#1}{\urlprefix }}%
\providecommand \urlprefix  [0]{URL }%
\providecommand \Eprint [0]{\href }%
\providecommand \doibase [0]{https://doi.org/}%
\providecommand \selectlanguage [0]{\@gobble}%
\providecommand \bibinfo  [0]{\@secondoftwo}%
\providecommand \bibfield  [0]{\@secondoftwo}%
\providecommand \translation [1]{[#1]}%
\providecommand \BibitemOpen [0]{}%
\providecommand \bibitemStop [0]{}%
\providecommand \bibitemNoStop [0]{.\EOS\space}%
\providecommand \EOS [0]{\spacefactor3000\relax}%
\providecommand \BibitemShut  [1]{\csname bibitem#1\endcsname}%
\let\auto@bib@innerbib\@empty
\bibitem [{\citenamefont {Foxen}\ \emph {et~al.}(2017)\citenamefont {Foxen},
  \citenamefont {Mutus}, \citenamefont {Lucero}, \citenamefont {Graff},
  \citenamefont {Megrant}, \citenamefont {Chen}, \citenamefont {Quintana},
  \citenamefont {Burkett}, \citenamefont {Kelly}, \citenamefont {Jeffrey},\
  and\ \citenamefont {et~al.}}]{Foxen_2017}%
  \BibitemOpen
  \bibfield  {author} {\bibinfo {author} {\bibfnamefont {B.}~\bibnamefont
  {Foxen}}, \bibinfo {author} {\bibfnamefont {J.~Y.}\ \bibnamefont {Mutus}},
  \bibinfo {author} {\bibfnamefont {E.}~\bibnamefont {Lucero}}, \bibinfo
  {author} {\bibfnamefont {R.}~\bibnamefont {Graff}}, \bibinfo {author}
  {\bibfnamefont {A.}~\bibnamefont {Megrant}}, \bibinfo {author} {\bibfnamefont
  {Y.}~\bibnamefont {Chen}}, \bibinfo {author} {\bibfnamefont {C.}~\bibnamefont
  {Quintana}}, \bibinfo {author} {\bibfnamefont {B.}~\bibnamefont {Burkett}},
  \bibinfo {author} {\bibfnamefont {J.}~\bibnamefont {Kelly}}, \bibinfo
  {author} {\bibfnamefont {E.}~\bibnamefont {Jeffrey}},\ and\ \bibinfo {author}
  {\bibnamefont {et~al.}},\ }\bibfield  {title} {\bibinfo {title} {Qubit
  compatible superconducting interconnects},\ }\href
  {https://doi.org/10.1088/2058-9565/aa94fc} {\bibfield  {journal} {\bibinfo
  {journal} {Quantum Science and Technology}\ }\textbf {\bibinfo {volume}
  {3}},\ \bibinfo {pages} {014005} (\bibinfo {year} {2017})}\BibitemShut
  {NoStop}%
\bibitem [{\citenamefont {Rosenberg}\ \emph {et~al.}(2017)\citenamefont
  {Rosenberg}, \citenamefont {Kim}, \citenamefont {Das}, \citenamefont {Yost},
  \citenamefont {Gustavsson}, \citenamefont {Hover}, \citenamefont {Krantz},
  \citenamefont {Melville}, \citenamefont {Racz}, \citenamefont {Samach},\ and\
  \citenamefont {et~al.}}]{Rosenberg_2017}%
  \BibitemOpen
  \bibfield  {author} {\bibinfo {author} {\bibfnamefont {D.}~\bibnamefont
  {Rosenberg}}, \bibinfo {author} {\bibfnamefont {D.}~\bibnamefont {Kim}},
  \bibinfo {author} {\bibfnamefont {R.}~\bibnamefont {Das}}, \bibinfo {author}
  {\bibfnamefont {D.}~\bibnamefont {Yost}}, \bibinfo {author} {\bibfnamefont
  {S.}~\bibnamefont {Gustavsson}}, \bibinfo {author} {\bibfnamefont
  {D.}~\bibnamefont {Hover}}, \bibinfo {author} {\bibfnamefont
  {P.}~\bibnamefont {Krantz}}, \bibinfo {author} {\bibfnamefont
  {A.}~\bibnamefont {Melville}}, \bibinfo {author} {\bibfnamefont
  {L.}~\bibnamefont {Racz}}, \bibinfo {author} {\bibfnamefont {G.~O.}\
  \bibnamefont {Samach}},\ and\ \bibinfo {author} {\bibnamefont {et~al.}},\
  }\bibfield  {title} {\bibinfo {title} {3{D} integrated superconducting
  qubits},\ }\href {https://doi.org/10.1038/s41534-017-0044-0} {\bibfield
  {journal} {\bibinfo  {journal} {npj Quantum Information}\ }\textbf {\bibinfo
  {volume} {3}},\ \bibinfo {pages} {42} (\bibinfo {year} {2017})}\BibitemShut
  {NoStop}%
\bibitem [{\citenamefont {Conner}\ \emph {et~al.}(2021)\citenamefont {Conner},
  \citenamefont {Bienfait}, \citenamefont {Chang}, \citenamefont {Chou},
  \citenamefont {Dumur}, \citenamefont {Grebel}, \citenamefont {Peairs},
  \citenamefont {Povey}, \citenamefont {Yan}, \citenamefont {Zhong},\ and\
  \citenamefont {et~al.}}]{Conner_2021}%
  \BibitemOpen
  \bibfield  {author} {\bibinfo {author} {\bibfnamefont {C.~R.}\ \bibnamefont
  {Conner}}, \bibinfo {author} {\bibfnamefont {A.}~\bibnamefont {Bienfait}},
  \bibinfo {author} {\bibfnamefont {H.-S.}\ \bibnamefont {Chang}}, \bibinfo
  {author} {\bibfnamefont {M.-H.}\ \bibnamefont {Chou}}, \bibinfo {author}
  {\bibnamefont {Dumur}}, \bibinfo {author} {\bibfnamefont {J.}~\bibnamefont
  {Grebel}}, \bibinfo {author} {\bibfnamefont {G.~A.}\ \bibnamefont {Peairs}},
  \bibinfo {author} {\bibfnamefont {R.~G.}\ \bibnamefont {Povey}}, \bibinfo
  {author} {\bibfnamefont {H.}~\bibnamefont {Yan}}, \bibinfo {author}
  {\bibfnamefont {Y.~P.}\ \bibnamefont {Zhong}},\ and\ \bibinfo {author}
  {\bibnamefont {et~al.}},\ }\bibfield  {title} {\bibinfo {title}
  {Superconducting qubits in a flip-chip architecture},\ }\href
  {https://doi.org/10.1063/5.0050173} {\bibfield  {journal} {\bibinfo
  {journal} {Applied Physics Letters}\ }\textbf {\bibinfo {volume} {118}},\
  \bibinfo {pages} {232602} (\bibinfo {year} {2021})}\BibitemShut {NoStop}%
\bibitem [{\citenamefont {Kosen}\ \emph {et~al.}(2022)\citenamefont {Kosen},
  \citenamefont {Li}, \citenamefont {Rommel}, \citenamefont {Shiri},
  \citenamefont {Warren}, \citenamefont {Grönberg}, \citenamefont {Salonen},
  \citenamefont {Abad}, \citenamefont {Biznárová}, \citenamefont {Caputo},\
  and\ \citenamefont {et~al.}}]{Kosen_2022}%
  \BibitemOpen
  \bibfield  {author} {\bibinfo {author} {\bibfnamefont {S.}~\bibnamefont
  {Kosen}}, \bibinfo {author} {\bibfnamefont {H.-X.}\ \bibnamefont {Li}},
  \bibinfo {author} {\bibfnamefont {M.}~\bibnamefont {Rommel}}, \bibinfo
  {author} {\bibfnamefont {D.}~\bibnamefont {Shiri}}, \bibinfo {author}
  {\bibfnamefont {C.}~\bibnamefont {Warren}}, \bibinfo {author} {\bibfnamefont
  {L.}~\bibnamefont {Grönberg}}, \bibinfo {author} {\bibfnamefont
  {J.}~\bibnamefont {Salonen}}, \bibinfo {author} {\bibfnamefont
  {T.}~\bibnamefont {Abad}}, \bibinfo {author} {\bibfnamefont {J.}~\bibnamefont
  {Biznárová}}, \bibinfo {author} {\bibfnamefont {M.}~\bibnamefont
  {Caputo}},\ and\ \bibinfo {author} {\bibnamefont {et~al.}},\ }\bibfield
  {title} {\bibinfo {title} {Building blocks of a flip-chip integrated
  superconducting quantum processor},\ }\href
  {https://doi.org/10.1088/2058-9565/ac734b} {\bibfield  {journal} {\bibinfo
  {journal} {Quantum Science and Technology}\ }\textbf {\bibinfo {volume}
  {7}},\ \bibinfo {pages} {035018} (\bibinfo {year} {2022})}\BibitemShut
  {NoStop}%
\bibitem [{\citenamefont {Norris}\ \emph {et~al.}(2024)\citenamefont {Norris},
  \citenamefont {Michaud}, \citenamefont {Pahl}, \citenamefont {Kerschbaum},
  \citenamefont {Eichler}, \citenamefont {Besse},\ and\ \citenamefont
  {Wallraff}}]{Norris_Michaud_Pahl_Kerschbaum_Eichler_Besse_Wallraff_2024}%
  \BibitemOpen
  \bibfield  {author} {\bibinfo {author} {\bibfnamefont {G.~J.}\ \bibnamefont
  {Norris}}, \bibinfo {author} {\bibfnamefont {L.}~\bibnamefont {Michaud}},
  \bibinfo {author} {\bibfnamefont {D.}~\bibnamefont {Pahl}}, \bibinfo {author}
  {\bibfnamefont {M.}~\bibnamefont {Kerschbaum}}, \bibinfo {author}
  {\bibfnamefont {C.}~\bibnamefont {Eichler}}, \bibinfo {author} {\bibfnamefont
  {J.-C.}\ \bibnamefont {Besse}},\ and\ \bibinfo {author} {\bibfnamefont
  {A.}~\bibnamefont {Wallraff}},\ }\bibfield  {title} {\bibinfo {title}
  {Improved parameter targeting in 3{D}-integrated superconducting circuits
  through a polymer spacer process},\ }\href
  {https://doi.org/10.1140/epjqt/s40507-023-00213-x} {\bibfield  {journal}
  {\bibinfo  {journal} {EPJ Quantum Technology}\ }\textbf {\bibinfo {volume}
  {11}},\ \bibinfo {pages} {5} (\bibinfo {year} {2024})}\BibitemShut {NoStop}%
\bibitem [{\citenamefont {DeNigris}\ \emph {et~al.}(2018)\citenamefont
  {DeNigris}, \citenamefont {Chervenak}, \citenamefont {Bandler}, \citenamefont
  {Chang}, \citenamefont {Costen}, \citenamefont {Eckart}, \citenamefont {Ha},
  \citenamefont {Kilbourne},\ and\ \citenamefont
  {Smith}}]{DeNigris_Chervenak_Bandler_Chang_Costen_Eckart_Ha_Kilbourne_Smith_2018}%
  \BibitemOpen
  \bibfield  {author} {\bibinfo {author} {\bibfnamefont {N.~S.}\ \bibnamefont
  {DeNigris}}, \bibinfo {author} {\bibfnamefont {J.~A.}\ \bibnamefont
  {Chervenak}}, \bibinfo {author} {\bibfnamefont {S.~R.}\ \bibnamefont
  {Bandler}}, \bibinfo {author} {\bibfnamefont {M.~P.}\ \bibnamefont {Chang}},
  \bibinfo {author} {\bibfnamefont {N.~P.}\ \bibnamefont {Costen}}, \bibinfo
  {author} {\bibfnamefont {M.~E.}\ \bibnamefont {Eckart}}, \bibinfo {author}
  {\bibfnamefont {J.~Y.}\ \bibnamefont {Ha}}, \bibinfo {author} {\bibfnamefont
  {C.~A.}\ \bibnamefont {Kilbourne}},\ and\ \bibinfo {author} {\bibfnamefont
  {S.~J.}\ \bibnamefont {Smith}},\ }\bibfield  {title} {\bibinfo {title}
  {Fabrication of flexible superconducting wiring with high current-carrying
  capacity indium interconnects},\ }\href
  {https://doi.org/10.1007/s10909-018-2019-8} {\bibfield  {journal} {\bibinfo
  {journal} {Journal of Low Temperature Physics}\ }\textbf {\bibinfo {volume}
  {193}},\ \bibinfo {pages} {687–694} (\bibinfo {year} {2018})}\BibitemShut
  {NoStop}%
\bibitem [{\citenamefont {Lucas}\ \emph {et~al.}(2022)\citenamefont {Lucas},
  \citenamefont {Biesecker}, \citenamefont {Doriese}, \citenamefont {Duff},
  \citenamefont {Hilton}, \citenamefont {Ullom}, \citenamefont {Vissers},\ and\
  \citenamefont
  {Schmidt}}]{Lucas_Biesecker_Doriese_Duff_Hilton_Ullom_Vissers_Schmidt_2022}%
  \BibitemOpen
  \bibfield  {author} {\bibinfo {author} {\bibfnamefont {T.~J.}\ \bibnamefont
  {Lucas}}, \bibinfo {author} {\bibfnamefont {J.~P.}\ \bibnamefont
  {Biesecker}}, \bibinfo {author} {\bibfnamefont {W.~B.}\ \bibnamefont
  {Doriese}}, \bibinfo {author} {\bibfnamefont {S.~M.}\ \bibnamefont {Duff}},
  \bibinfo {author} {\bibfnamefont {G.~C.}\ \bibnamefont {Hilton}}, \bibinfo
  {author} {\bibfnamefont {J.~N.}\ \bibnamefont {Ullom}}, \bibinfo {author}
  {\bibfnamefont {M.~R.}\ \bibnamefont {Vissers}},\ and\ \bibinfo {author}
  {\bibfnamefont {D.~R.}\ \bibnamefont {Schmidt}},\ }\bibfield  {title}
  {\bibinfo {title} {Indium bump process for low-temperature detectors and
  readout},\ }\href {https://doi.org/10.1007/s10909-022-02728-6} {\bibfield
  {journal} {\bibinfo  {journal} {Journal of Low Temperature Physics}\ }\textbf
  {\bibinfo {volume} {209}},\ \bibinfo {pages} {293–298} (\bibinfo {year}
  {2022})}\BibitemShut {NoStop}%
\bibitem [{\citenamefont {Plötner}\ \emph {et~al.}(1991)\citenamefont
  {Plötner}, \citenamefont {Sadowski}, \citenamefont {Rzepka},\ and\
  \citenamefont {Blasek}}]{Plötner_Sadowski_Rzepka_Blasek_1991}%
  \BibitemOpen
  \bibfield  {author} {\bibinfo {author} {\bibfnamefont {M.}~\bibnamefont
  {Plötner}}, \bibinfo {author} {\bibfnamefont {G.}~\bibnamefont {Sadowski}},
  \bibinfo {author} {\bibfnamefont {S.}~\bibnamefont {Rzepka}},\ and\ \bibinfo
  {author} {\bibfnamefont {G.}~\bibnamefont {Blasek}},\ }\bibfield  {title}
  {\bibinfo {title} {Aspects of indium solder bumping and indium bump bonding
  useful for assembling cooled mosaic sensors},\ }\href
  {https://doi.org/10.1108/eb044447} {\bibfield  {journal} {\bibinfo  {journal}
  {Microelectronics International}\ }\textbf {\bibinfo {volume} {8}},\ \bibinfo
  {pages} {27–30} (\bibinfo {year} {1991})}\BibitemShut {NoStop}%
\bibitem [{\citenamefont {Huang}\ \emph {et~al.}(2010)\citenamefont {Huang},
  \citenamefont {Xu}, \citenamefont {Yuan}, \citenamefont {Cheng},\ and\
  \citenamefont {Luo}}]{Huang_Xu_Yuan_Cheng_Luo_2010}%
  \BibitemOpen
  \bibfield  {author} {\bibinfo {author} {\bibfnamefont {Q.}~\bibnamefont
  {Huang}}, \bibinfo {author} {\bibfnamefont {G.}~\bibnamefont {Xu}}, \bibinfo
  {author} {\bibfnamefont {Y.}~\bibnamefont {Yuan}}, \bibinfo {author}
  {\bibfnamefont {X.}~\bibnamefont {Cheng}},\ and\ \bibinfo {author}
  {\bibfnamefont {L.}~\bibnamefont {Luo}},\ }\bibfield  {title} {\bibinfo
  {title} {Development of indium bumping technology through az9260 resist
  electroplating},\ }\href {https://doi.org/10.1088/0960-1317/20/5/055035}
  {\bibfield  {journal} {\bibinfo  {journal} {Journal of Micromechanics and
  Microengineering}\ }\textbf {\bibinfo {volume} {20}},\ \bibinfo {pages}
  {055035} (\bibinfo {year} {2010})}\BibitemShut {NoStop}%
\bibitem [{\citenamefont {Wade}\ and\ \citenamefont
  {Banister}(1973)}]{Wade1973}%
  \BibitemOpen
  \bibfield  {author} {\bibinfo {author} {\bibfnamefont {K.}~\bibnamefont
  {Wade}}\ and\ \bibinfo {author} {\bibfnamefont {A.}~\bibnamefont
  {Banister}},\ }\href {https://books.google.se/books?id=QwNPDAAAQBAJ} {\emph
  {\bibinfo {title} {The Chemistry of Aluminium, Gallium, Indium and
  Thallium}}},\ \bibinfo {edition} {1st}\ ed.,\ \bibinfo {series} {Pergamon
  Texts in Inorganic Chemistry}, Vol.~\bibinfo {volume} {12}\ (\bibinfo
  {publisher} {Pergamon Press},\ \bibinfo {address} {Oxford, UK},\ \bibinfo
  {year} {1973})\ p.\ \bibinfo {pages} {107}\BibitemShut {NoStop}%
\bibitem [{\citenamefont {Gordon}\ \emph {et~al.}(1999)\citenamefont {Gordon},
  \citenamefont {Liu}, \citenamefont {Broomhall-Dillard},\ and\ \citenamefont
  {Shi}}]{Gordon_1999}%
  \BibitemOpen
  \bibfield  {author} {\bibinfo {author} {\bibfnamefont {R.~G.}\ \bibnamefont
  {Gordon}}, \bibinfo {author} {\bibfnamefont {X.}~\bibnamefont {Liu}},
  \bibinfo {author} {\bibfnamefont {R.~N.}\ \bibnamefont {Broomhall-Dillard}},\
  and\ \bibinfo {author} {\bibfnamefont {Y.}~\bibnamefont {Shi}},\ }\bibfield
  {title} {\bibinfo {title} {Highly conformal diffusion barriers of amorphous
  niobium nitride},\ }\href {https://doi.org/10.1557/proc-564-335} {\bibfield
  {journal} {\bibinfo  {journal} {MRS Proceedings}\ }\textbf {\bibinfo {volume}
  {564}},\ \bibinfo {pages} {335} (\bibinfo {year} {1999})}\BibitemShut
  {NoStop}%
\bibitem [{\citenamefont {Thomas}\ \emph {et~al.}(2022)\citenamefont {Thomas},
  \citenamefont {Michel}, \citenamefont {Deschaseaux}, \citenamefont
  {Charbonnier}, \citenamefont {Souil}, \citenamefont {Vermande}, \citenamefont
  {Campo}, \citenamefont {Farjot}, \citenamefont {Rodriguez}, \citenamefont
  {Romano},\ and\ \citenamefont
  {et~al.}}]{Thomas_Michel_Deschaseaux_Charbonnier_Souil_Vermande_Campo_Farjot_Rodriguez_Romano_et_al._2022}%
  \BibitemOpen
  \bibfield  {author} {\bibinfo {author} {\bibfnamefont {C.}~\bibnamefont
  {Thomas}}, \bibinfo {author} {\bibfnamefont {J.-P.}\ \bibnamefont {Michel}},
  \bibinfo {author} {\bibfnamefont {E.}~\bibnamefont {Deschaseaux}}, \bibinfo
  {author} {\bibfnamefont {J.}~\bibnamefont {Charbonnier}}, \bibinfo {author}
  {\bibfnamefont {R.}~\bibnamefont {Souil}}, \bibinfo {author} {\bibfnamefont
  {E.}~\bibnamefont {Vermande}}, \bibinfo {author} {\bibfnamefont
  {A.}~\bibnamefont {Campo}}, \bibinfo {author} {\bibfnamefont
  {T.}~\bibnamefont {Farjot}}, \bibinfo {author} {\bibfnamefont
  {G.}~\bibnamefont {Rodriguez}}, \bibinfo {author} {\bibfnamefont
  {G.}~\bibnamefont {Romano}},\ and\ \bibinfo {author} {\bibnamefont
  {et~al.}},\ }\bibfield  {title} {\bibinfo {title} {Superconducting routing
  platform for large-scale integration of quantum technologies},\ }\href
  {https://doi.org/10.1088/2633-4356/ac88ae} {\bibfield  {journal} {\bibinfo
  {journal} {Materials for Quantum Technology}\ }\textbf {\bibinfo {volume}
  {2}},\ \bibinfo {pages} {035001} (\bibinfo {year} {2022})}\BibitemShut
  {NoStop}%
\bibitem [{\citenamefont {Volpert}\ \emph {et~al.}(2010)\citenamefont
  {Volpert}, \citenamefont {Roulet}, \citenamefont {Boronat}, \citenamefont
  {Borel}, \citenamefont {Pocas},\ and\ \citenamefont
  {Ribot}}]{Volpert2010IndiumDP}%
  \BibitemOpen
  \bibfield  {author} {\bibinfo {author} {\bibfnamefont {M.}~\bibnamefont
  {Volpert}}, \bibinfo {author} {\bibfnamefont {L.}~\bibnamefont {Roulet}},
  \bibinfo {author} {\bibfnamefont {J.-F.}\ \bibnamefont {Boronat}}, \bibinfo
  {author} {\bibfnamefont {I.}~\bibnamefont {Borel}}, \bibinfo {author}
  {\bibfnamefont {S.}~\bibnamefont {Pocas}},\ and\ \bibinfo {author}
  {\bibfnamefont {H.}~\bibnamefont {Ribot}},\ }\bibfield  {title} {\bibinfo
  {title} {Indium deposition processes for ultra fine pitch 3{D}
  interconnections},\ }\href {https://doi.org/10.1109/ECTC.2010.5490736}
  {\bibfield  {journal} {\bibinfo  {journal} {2010 Proceedings 60th Electronic
  Components and Technology Conference (ECTC)}\ ,\ \bibinfo {pages} {1739}}
  (\bibinfo {year} {2010})}\BibitemShut {NoStop}%
\bibitem [{\citenamefont {Kozłowski}\ \emph {et~al.}(2021)\citenamefont
  {Kozłowski}, \citenamefont {Czuba}, \citenamefont {Chmielewski},
  \citenamefont {Ratajczak}, \citenamefont {Branas}, \citenamefont {Korczyc},
  \citenamefont {Regiński},\ and\ \citenamefont
  {Jasik}}]{Kozłowski_Czuba_Chmielewski_Ratajczak_Branas_Korczyc_Regiński_Jasik_2021}%
  \BibitemOpen
  \bibfield  {author} {\bibinfo {author} {\bibfnamefont {P.}~\bibnamefont
  {Kozłowski}}, \bibinfo {author} {\bibfnamefont {K.}~\bibnamefont {Czuba}},
  \bibinfo {author} {\bibfnamefont {K.}~\bibnamefont {Chmielewski}}, \bibinfo
  {author} {\bibfnamefont {J.}~\bibnamefont {Ratajczak}}, \bibinfo {author}
  {\bibfnamefont {J.}~\bibnamefont {Branas}}, \bibinfo {author} {\bibfnamefont
  {A.}~\bibnamefont {Korczyc}}, \bibinfo {author} {\bibfnamefont
  {K.}~\bibnamefont {Regiński}},\ and\ \bibinfo {author} {\bibfnamefont
  {A.}~\bibnamefont {Jasik}},\ }\bibfield  {title} {\bibinfo {title}
  {Indium-based micro-bump array fabrication technology with added pre-reflow
  wet etching and annealing},\ }\href {https://doi.org/10.3390/ma14216269}
  {\bibfield  {journal} {\bibinfo  {journal} {Materials}\ }\textbf {\bibinfo
  {volume} {14}},\ \bibinfo {pages} {6269} (\bibinfo {year}
  {2021})}\BibitemShut {NoStop}%
\bibitem [{\citenamefont {Das}\ \emph {et~al.}(2019)\citenamefont {Das},
  \citenamefont {Bolkhovsky}, \citenamefont {Galbraith}, \citenamefont {Oates},
  \citenamefont {Plant}, \citenamefont {Lambert}, \citenamefont {Zarr},
  \citenamefont {Rastogi}, \citenamefont {Shapiro}, \citenamefont {Docanto},
  \citenamefont {Weir},\ and\ \citenamefont {Johnson}}]{Das_2019}%
  \BibitemOpen
  \bibfield  {author} {\bibinfo {author} {\bibfnamefont {R.}~\bibnamefont
  {Das}}, \bibinfo {author} {\bibfnamefont {V.}~\bibnamefont {Bolkhovsky}},
  \bibinfo {author} {\bibfnamefont {C.}~\bibnamefont {Galbraith}}, \bibinfo
  {author} {\bibfnamefont {D.}~\bibnamefont {Oates}}, \bibinfo {author}
  {\bibfnamefont {J.}~\bibnamefont {Plant}}, \bibinfo {author} {\bibfnamefont
  {R.}~\bibnamefont {Lambert}}, \bibinfo {author} {\bibfnamefont
  {S.}~\bibnamefont {Zarr}}, \bibinfo {author} {\bibfnamefont {R.}~\bibnamefont
  {Rastogi}}, \bibinfo {author} {\bibfnamefont {D.}~\bibnamefont {Shapiro}},
  \bibinfo {author} {\bibfnamefont {M.}~\bibnamefont {Docanto}}, \bibinfo
  {author} {\bibfnamefont {T.}~\bibnamefont {Weir}},\ and\ \bibinfo {author}
  {\bibfnamefont {L.}~\bibnamefont {Johnson}},\ }\bibfield  {title} {\bibinfo
  {title} {Interconnect scheme for die-to-die and die-to-wafer-level
  heterogeneous integration for high-performance computing},\ }in\ \href
  {https://doi.org/10.1109/ECTC.2019.00248} {\emph {\bibinfo {booktitle} {2019
  IEEE 69th Electronic Components and Technology Conference (ECTC)}}}\
  (\bibinfo {year} {2019})\ pp.\ \bibinfo {pages} {1611--1621}\BibitemShut
  {NoStop}%
\bibitem [{\citenamefont {Satzinger}\ \emph {et~al.}(2019)\citenamefont
  {Satzinger}, \citenamefont {Conner}, \citenamefont {Bienfait}, \citenamefont
  {Chang}, \citenamefont {Chou}, \citenamefont {Cleland}, \citenamefont
  {Dumur}, \citenamefont {Grebel}, \citenamefont {Peairs}, \citenamefont
  {Povey},\ and\ \citenamefont
  {et~al.}}]{Satzinger_Conner_Bienfait_Chang_Chou_Cleland_Dumur_Grebel_Peairs_Povey_et_al._2019}%
  \BibitemOpen
  \bibfield  {author} {\bibinfo {author} {\bibfnamefont {K.~J.}\ \bibnamefont
  {Satzinger}}, \bibinfo {author} {\bibfnamefont {C.~R.}\ \bibnamefont
  {Conner}}, \bibinfo {author} {\bibfnamefont {A.}~\bibnamefont {Bienfait}},
  \bibinfo {author} {\bibfnamefont {H.-S.}\ \bibnamefont {Chang}}, \bibinfo
  {author} {\bibfnamefont {M.-H.}\ \bibnamefont {Chou}}, \bibinfo {author}
  {\bibfnamefont {A.~Y.}\ \bibnamefont {Cleland}}, \bibinfo {author}
  {\bibnamefont {Dumur}}, \bibinfo {author} {\bibfnamefont {J.}~\bibnamefont
  {Grebel}}, \bibinfo {author} {\bibfnamefont {G.~A.}\ \bibnamefont {Peairs}},
  \bibinfo {author} {\bibfnamefont {R.~G.}\ \bibnamefont {Povey}},\ and\
  \bibinfo {author} {\bibnamefont {et~al.}},\ }\bibfield  {title} {\bibinfo
  {title} {Simple non-galvanic flip-chip integration method for hybrid quantum
  systems},\ }\href {https://doi.org/10.1063/1.5089888} {\bibfield  {journal}
  {\bibinfo  {journal} {Applied Physics Letters}\ }\textbf {\bibinfo {volume}
  {114}},\ \bibinfo {pages} {173501} (\bibinfo {year} {2019})}\BibitemShut
  {NoStop}%
\bibitem [{\citenamefont {CAPLINQ}()}]{Caplinq}%
  \BibitemOpen
  \bibfield  {author} {\bibinfo {author} {\bibnamefont {CAPLINQ}},\ }\href
  {https://www.caplinq.com/solder-spheres.html} {\bibinfo {title} {In100 pure
  indium solder spheres}}\BibitemShut {NoStop}%
\bibitem [{\citenamefont {Zhao}\ \emph {et~al.}(2020)\citenamefont {Zhao},
  \citenamefont {Wang}, \citenamefont {Ryu}, \citenamefont {Frisenda},\ and\
  \citenamefont {Castellanos-Gomez}}]{Zhao_2020}%
  \BibitemOpen
  \bibfield  {author} {\bibinfo {author} {\bibfnamefont {Q.}~\bibnamefont
  {Zhao}}, \bibinfo {author} {\bibfnamefont {T.}~\bibnamefont {Wang}}, \bibinfo
  {author} {\bibfnamefont {Y.~K.}\ \bibnamefont {Ryu}}, \bibinfo {author}
  {\bibfnamefont {R.}~\bibnamefont {Frisenda}},\ and\ \bibinfo {author}
  {\bibfnamefont {A.}~\bibnamefont {Castellanos-Gomez}},\ }\bibfield  {title}
  {\bibinfo {title} {An inexpensive system for the deterministic transfer of
  2{D} materials},\ }\href {https://doi.org/10.1088/2515-7639/ab6a72}
  {\bibfield  {journal} {\bibinfo  {journal} {Journal of Physics: Materials}\
  }\textbf {\bibinfo {volume} {3}},\ \bibinfo {pages} {016001} (\bibinfo {year}
  {2020})}\BibitemShut {NoStop}%
\bibitem [{\citenamefont {Latorre}\ \emph {et~al.}(2022)\citenamefont
  {Latorre}, \citenamefont {Paradkar}, \citenamefont {Hambraeus}, \citenamefont
  {Higgins},\ and\ \citenamefont {Wieczorek}}]{martiIEEE2022}%
  \BibitemOpen
  \bibfield  {author} {\bibinfo {author} {\bibfnamefont {M.~G.}\ \bibnamefont
  {Latorre}}, \bibinfo {author} {\bibfnamefont {A.}~\bibnamefont {Paradkar}},
  \bibinfo {author} {\bibfnamefont {D.}~\bibnamefont {Hambraeus}}, \bibinfo
  {author} {\bibfnamefont {G.}~\bibnamefont {Higgins}},\ and\ \bibinfo {author}
  {\bibfnamefont {W.}~\bibnamefont {Wieczorek}},\ }\bibfield  {title} {\bibinfo
  {title} {A chip-based superconducting magnetic trap for levitating
  superconducting microparticles},\ }\href
  {https://doi.org/10.1109/TASC.2022.3147730} {\bibfield  {journal} {\bibinfo
  {journal} {IEEE Transactions on Applied Superconductivity}\ }\textbf
  {\bibinfo {volume} {32}},\ \bibinfo {pages} {1} (\bibinfo {year}
  {2022})}\BibitemShut {NoStop}%
\bibitem [{\citenamefont {Latorre}\ \emph {et~al.}(2023)\citenamefont
  {Latorre}, \citenamefont {Higgins}, \citenamefont {Paradkar}, \citenamefont
  {Bauch},\ and\ \citenamefont {Wieczorek}}]{martiPRA2023}%
  \BibitemOpen
  \bibfield  {author} {\bibinfo {author} {\bibfnamefont {M.~G.}\ \bibnamefont
  {Latorre}}, \bibinfo {author} {\bibfnamefont {G.}~\bibnamefont {Higgins}},
  \bibinfo {author} {\bibfnamefont {A.}~\bibnamefont {Paradkar}}, \bibinfo
  {author} {\bibfnamefont {T.}~\bibnamefont {Bauch}},\ and\ \bibinfo {author}
  {\bibfnamefont {W.}~\bibnamefont {Wieczorek}},\ }\bibfield  {title} {\bibinfo
  {title} {Superconducting microsphere magnetically levitated in an anharmonic
  potential with integrated magnetic readout},\ }\href
  {https://doi.org/10.1103/PhysRevApplied.19.054047} {\bibfield  {journal}
  {\bibinfo  {journal} {Phys. Rev. Appl.}\ }\textbf {\bibinfo {volume} {19}},\
  \bibinfo {pages} {054047} (\bibinfo {year} {2023})}\BibitemShut {NoStop}%
\bibitem [{\citenamefont {Graninger}\ \emph {et~al.}(2019)\citenamefont
  {Graninger}, \citenamefont {Longo}, \citenamefont {Rennie}, \citenamefont
  {Shiskowski}, \citenamefont {Mercurio}, \citenamefont {Lilly},\ and\
  \citenamefont
  {Smith}}]{Graninger_Longo_Rennie_Shiskowski_Mercurio_Lilly_Smith_2019}%
  \BibitemOpen
  \bibfield  {author} {\bibinfo {author} {\bibfnamefont {A.~L.}\ \bibnamefont
  {Graninger}}, \bibinfo {author} {\bibfnamefont {M.~R.}\ \bibnamefont
  {Longo}}, \bibinfo {author} {\bibfnamefont {M.}~\bibnamefont {Rennie}},
  \bibinfo {author} {\bibfnamefont {R.}~\bibnamefont {Shiskowski}}, \bibinfo
  {author} {\bibfnamefont {K.}~\bibnamefont {Mercurio}}, \bibinfo {author}
  {\bibfnamefont {M.}~\bibnamefont {Lilly}},\ and\ \bibinfo {author}
  {\bibfnamefont {C.~H.}\ \bibnamefont {Smith}},\ }\bibfield  {title} {\bibinfo
  {title} {Superconducting on-chip solenoid for josephson junction
  characterization},\ }\href {https://doi.org/10.1063/1.5110170} {\bibfield
  {journal} {\bibinfo  {journal} {Applied Physics Letters}\ }\textbf {\bibinfo
  {volume} {115}},\ \bibinfo {pages} {032601} (\bibinfo {year}
  {2019})}\BibitemShut {NoStop}%
\bibitem [{\citenamefont {Verjauw}\ \emph {et~al.}(2021)\citenamefont
  {Verjauw}, \citenamefont {Potocnik}, \citenamefont {Mongillo}, \citenamefont
  {Acharya}, \citenamefont {Mohiyaddin}, \citenamefont {Simion}, \citenamefont
  {Pacco}, \citenamefont {Ivanov}, \citenamefont {Wan}, \citenamefont
  {Vanleenhove}, \citenamefont {Souriau}, \citenamefont {Jussot}, \citenamefont
  {Thiam}, \citenamefont {Swerts}, \citenamefont {Piao}, \citenamefont {Couet},
  \citenamefont {Heyns}, \citenamefont {Govoreanu},\ and\ \citenamefont
  {Radu}}]{Verjauw_2021}%
  \BibitemOpen
  \bibfield  {author} {\bibinfo {author} {\bibfnamefont {J.}~\bibnamefont
  {Verjauw}}, \bibinfo {author} {\bibfnamefont {A.}~\bibnamefont {Potocnik}},
  \bibinfo {author} {\bibfnamefont {M.}~\bibnamefont {Mongillo}}, \bibinfo
  {author} {\bibfnamefont {R.}~\bibnamefont {Acharya}}, \bibinfo {author}
  {\bibfnamefont {F.}~\bibnamefont {Mohiyaddin}}, \bibinfo {author}
  {\bibfnamefont {G.}~\bibnamefont {Simion}}, \bibinfo {author} {\bibfnamefont
  {A.}~\bibnamefont {Pacco}}, \bibinfo {author} {\bibfnamefont
  {T.}~\bibnamefont {Ivanov}}, \bibinfo {author} {\bibfnamefont
  {D.}~\bibnamefont {Wan}}, \bibinfo {author} {\bibfnamefont {A.}~\bibnamefont
  {Vanleenhove}}, \bibinfo {author} {\bibfnamefont {L.}~\bibnamefont
  {Souriau}}, \bibinfo {author} {\bibfnamefont {J.}~\bibnamefont {Jussot}},
  \bibinfo {author} {\bibfnamefont {A.}~\bibnamefont {Thiam}}, \bibinfo
  {author} {\bibfnamefont {J.}~\bibnamefont {Swerts}}, \bibinfo {author}
  {\bibfnamefont {X.}~\bibnamefont {Piao}}, \bibinfo {author} {\bibfnamefont
  {S.}~\bibnamefont {Couet}}, \bibinfo {author} {\bibfnamefont
  {M.}~\bibnamefont {Heyns}}, \bibinfo {author} {\bibfnamefont
  {B.}~\bibnamefont {Govoreanu}},\ and\ \bibinfo {author} {\bibfnamefont
  {I.}~\bibnamefont {Radu}},\ }\bibfield  {title} {\bibinfo {title}
  {Investigation of microwave loss induced by oxide regrowth in high-{Q}
  niobium resonators},\ }\href
  {https://doi.org/10.1103/PhysRevApplied.16.014018} {\bibfield  {journal}
  {\bibinfo  {journal} {Phys. Rev. Appl.}\ }\textbf {\bibinfo {volume} {16}},\
  \bibinfo {pages} {014018} (\bibinfo {year} {2021})}\BibitemShut {NoStop}%
\bibitem [{\citenamefont {Alto\'e}\ \emph {et~al.}(2022)\citenamefont
  {Alto\'e}, \citenamefont {Banerjee}, \citenamefont {Berk}, \citenamefont
  {Hajr}, \citenamefont {Schwartzberg}, \citenamefont {Song}, \citenamefont
  {Alghadeer}, \citenamefont {Aloni}, \citenamefont {Elowson}, \citenamefont
  {Kreikebaum}, \citenamefont {Wong}, \citenamefont {Griffin}, \citenamefont
  {Rao}, \citenamefont {Weber-Bargioni}, \citenamefont {Minor}, \citenamefont
  {Santiago}, \citenamefont {Cabrini}, \citenamefont {Siddiqi},\ and\
  \citenamefont {Ogletree}}]{Virginia_2022}%
  \BibitemOpen
  \bibfield  {author} {\bibinfo {author} {\bibfnamefont {M.~V.~P.}\
  \bibnamefont {Alto\'e}}, \bibinfo {author} {\bibfnamefont {A.}~\bibnamefont
  {Banerjee}}, \bibinfo {author} {\bibfnamefont {C.}~\bibnamefont {Berk}},
  \bibinfo {author} {\bibfnamefont {A.}~\bibnamefont {Hajr}}, \bibinfo {author}
  {\bibfnamefont {A.}~\bibnamefont {Schwartzberg}}, \bibinfo {author}
  {\bibfnamefont {C.}~\bibnamefont {Song}}, \bibinfo {author} {\bibfnamefont
  {M.}~\bibnamefont {Alghadeer}}, \bibinfo {author} {\bibfnamefont
  {S.}~\bibnamefont {Aloni}}, \bibinfo {author} {\bibfnamefont {M.~J.}\
  \bibnamefont {Elowson}}, \bibinfo {author} {\bibfnamefont {J.~M.}\
  \bibnamefont {Kreikebaum}}, \bibinfo {author} {\bibfnamefont {E.~K.}\
  \bibnamefont {Wong}}, \bibinfo {author} {\bibfnamefont {S.~M.}\ \bibnamefont
  {Griffin}}, \bibinfo {author} {\bibfnamefont {S.}~\bibnamefont {Rao}},
  \bibinfo {author} {\bibfnamefont {A.}~\bibnamefont {Weber-Bargioni}},
  \bibinfo {author} {\bibfnamefont {A.~M.}\ \bibnamefont {Minor}}, \bibinfo
  {author} {\bibfnamefont {D.~I.}\ \bibnamefont {Santiago}}, \bibinfo {author}
  {\bibfnamefont {S.}~\bibnamefont {Cabrini}}, \bibinfo {author} {\bibfnamefont
  {I.}~\bibnamefont {Siddiqi}},\ and\ \bibinfo {author} {\bibfnamefont {D.~F.}\
  \bibnamefont {Ogletree}},\ }\bibfield  {title} {\bibinfo {title}
  {Localization and mitigation of loss in niobium superconducting circuits},\
  }\href {https://doi.org/10.1103/PRXQuantum.3.020312} {\bibfield  {journal}
  {\bibinfo  {journal} {PRX Quantum}\ }\textbf {\bibinfo {volume} {3}},\
  \bibinfo {pages} {020312} (\bibinfo {year} {2022})}\BibitemShut {NoStop}%
\bibitem [{\citenamefont {Das}\ \emph {et~al.}(2018)\citenamefont {Das},
  \citenamefont {Yoder}, \citenamefont {Rosenberg}, \citenamefont {Kim},
  \citenamefont {Yost}, \citenamefont {Mallek}, \citenamefont {Hover},
  \citenamefont {Bolkhovsky}, \citenamefont {Kerman},\ and\ \citenamefont
  {Oliver}}]{Das_2018}%
  \BibitemOpen
  \bibfield  {author} {\bibinfo {author} {\bibfnamefont {R.~N.}\ \bibnamefont
  {Das}}, \bibinfo {author} {\bibfnamefont {J.}~\bibnamefont {Yoder}}, \bibinfo
  {author} {\bibfnamefont {D.}~\bibnamefont {Rosenberg}}, \bibinfo {author}
  {\bibfnamefont {D.}~\bibnamefont {Kim}}, \bibinfo {author} {\bibfnamefont
  {D.}~\bibnamefont {Yost}}, \bibinfo {author} {\bibfnamefont {J.}~\bibnamefont
  {Mallek}}, \bibinfo {author} {\bibfnamefont {D.}~\bibnamefont {Hover}},
  \bibinfo {author} {\bibfnamefont {V.}~\bibnamefont {Bolkhovsky}}, \bibinfo
  {author} {\bibfnamefont {A.}~\bibnamefont {Kerman}},\ and\ \bibinfo {author}
  {\bibfnamefont {W.}~\bibnamefont {Oliver}},\ }\bibfield  {title} {\bibinfo
  {title} {Cryogenic qubit integration for quantum computing},\ }in\ \href
  {https://doi.org/10.1109/ECTC.2018.00080} {\emph {\bibinfo {booktitle} {2018
  IEEE 68th Electronic Components and Technology Conference (ECTC)}}}\
  (\bibinfo {year} {2018})\ pp.\ \bibinfo {pages} {504--514}\BibitemShut
  {NoStop}%
\bibitem [{\citenamefont {Hamilton}\ \emph {et~al.}(1965)\citenamefont
  {Hamilton}, \citenamefont {Raub}, \citenamefont {Matthias}, \citenamefont
  {Corenzwit},\ and\ \citenamefont
  {Hull}}]{Hamilton_Raub_Matthias_Corenzwit_Hull_1965a}%
  \BibitemOpen
  \bibfield  {author} {\bibinfo {author} {\bibfnamefont {D.}~\bibnamefont
  {Hamilton}}, \bibinfo {author} {\bibfnamefont {C.}~\bibnamefont {Raub}},
  \bibinfo {author} {\bibfnamefont {B.}~\bibnamefont {Matthias}}, \bibinfo
  {author} {\bibfnamefont {E.}~\bibnamefont {Corenzwit}},\ and\ \bibinfo
  {author} {\bibfnamefont {G.}~\bibnamefont {Hull}},\ }\bibfield  {title}
  {\bibinfo {title} {Some new superconducting compounds},\ }\href
  {https://doi.org/10.1016/0022-3697(65)90144-7} {\bibfield  {journal}
  {\bibinfo  {journal} {Journal of Physics and Chemistry of Solids}\ }\textbf
  {\bibinfo {volume} {26}},\ \bibinfo {pages} {665–667} (\bibinfo {year}
  {1965})}\BibitemShut {NoStop}%
\bibitem [{\citenamefont {Schoeller}\ and\ \citenamefont
  {Cho}(2009)}]{Schoeller2009}%
  \BibitemOpen
  \bibfield  {author} {\bibinfo {author} {\bibfnamefont {H.}~\bibnamefont
  {Schoeller}}\ and\ \bibinfo {author} {\bibfnamefont {J.}~\bibnamefont
  {Cho}},\ }\bibfield  {title} {\bibinfo {title} {Oxidation and reduction
  behavior of pure indium},\ }\href {https://doi.org/10.1557/JMR.2009.0040}
  {\bibfield  {journal} {\bibinfo  {journal} {Journal of Materials Research}\
  }\textbf {\bibinfo {volume} {24}},\ \bibinfo {pages} {386–393} (\bibinfo
  {year} {2009})}\BibitemShut {NoStop}%
\bibitem [{\citenamefont {Il’in}\ \emph {et~al.}(2008)\citenamefont
  {Il’in}, \citenamefont {Stockhausen}, \citenamefont {Siegel}, \citenamefont
  {Semenov}, \citenamefont {Richter},\ and\ \citenamefont
  {H{\"u}bers}}]{il2008nbn}%
  \BibitemOpen
  \bibfield  {author} {\bibinfo {author} {\bibfnamefont {K.~S.}\ \bibnamefont
  {Il’in}}, \bibinfo {author} {\bibfnamefont {A.}~\bibnamefont
  {Stockhausen}}, \bibinfo {author} {\bibfnamefont {M.}~\bibnamefont {Siegel}},
  \bibinfo {author} {\bibfnamefont {A.~D.}\ \bibnamefont {Semenov}}, \bibinfo
  {author} {\bibfnamefont {H.}~\bibnamefont {Richter}},\ and\ \bibinfo {author}
  {\bibfnamefont {H.-W.}\ \bibnamefont {H{\"u}bers}},\ }\bibfield  {title}
  {\bibinfo {title} {$\mathrm{Nb}$ {HEB} for {TH}z radiation: Technological
  issues and proximity effect},\ }in\ \href
  {http://www.nrao.edu/meetings/isstt/papers/2008/2008403408.pdf} {\emph
  {\bibinfo {booktitle} {Proceedings of the 19th International Symposium on
  Space Terahertz Technology, ISSTT 2008}}}\ (\bibinfo {year} {2008})\ pp.\
  \bibinfo {pages} {367--372}\BibitemShut {NoStop}%
\bibitem [{\citenamefont {Kim}\ \emph {et~al.}(2009)\citenamefont {Kim},
  \citenamefont {Kahng}, \citenamefont {Choi},\ and\ \citenamefont
  {Lee}}]{Kim_2009}%
  \BibitemOpen
  \bibfield  {author} {\bibinfo {author} {\bibfnamefont {Y.~W.}\ \bibnamefont
  {Kim}}, \bibinfo {author} {\bibfnamefont {Y.~H.}\ \bibnamefont {Kahng}},
  \bibinfo {author} {\bibfnamefont {J.-H.}\ \bibnamefont {Choi}},\ and\
  \bibinfo {author} {\bibfnamefont {S.-G.}\ \bibnamefont {Lee}},\ }\bibfield
  {title} {\bibinfo {title} {Critical properties of submicrometer-patterned
  $\mathrm{Nb}$ thin film},\ }\href {https://doi.org/10.1109/TASC.2009.2019099}
  {\bibfield  {journal} {\bibinfo  {journal} {IEEE Transactions on Applied
  Superconductivity}\ }\textbf {\bibinfo {volume} {19}},\ \bibinfo {pages}
  {2649} (\bibinfo {year} {2009})}\BibitemShut {NoStop}%
\bibitem [{\citenamefont {Niepce}(2020)}]{Niepce_2020}%
  \BibitemOpen
  \bibfield  {author} {\bibinfo {author} {\bibfnamefont {D.}~\bibnamefont
  {Niepce}},\ }\emph {\bibinfo {title} {Superinductance and fluctuating
  two-level systems loss and noise in disordered and non-disordered
  superconducting quantum devices}},\ \href
  {https://research.chalmers.se/en/publication/?id=518627} {Ph.D. thesis},\
  \bibinfo  {school} {Chalmers University of Technology} (\bibinfo {year}
  {2020})\BibitemShut {NoStop}%
\bibitem [{\citenamefont {Joshi}\ \emph {et~al.}(2023)\citenamefont {Joshi},
  \citenamefont {Ghimire}, \citenamefont {Tanatar}, \citenamefont {Datta},
  \citenamefont {Oh}, \citenamefont {Zhou}, \citenamefont {Kopas},
  \citenamefont {Marshall}, \citenamefont {Mutus}, \citenamefont {Slaughter},\
  and\ \citenamefont {et~al.}}]{Joshi_2023}%
  \BibitemOpen
  \bibfield  {author} {\bibinfo {author} {\bibfnamefont {K.~R.}\ \bibnamefont
  {Joshi}}, \bibinfo {author} {\bibfnamefont {S.}~\bibnamefont {Ghimire}},
  \bibinfo {author} {\bibfnamefont {M.~A.}\ \bibnamefont {Tanatar}}, \bibinfo
  {author} {\bibfnamefont {A.}~\bibnamefont {Datta}}, \bibinfo {author}
  {\bibfnamefont {J.-S.}\ \bibnamefont {Oh}}, \bibinfo {author} {\bibfnamefont
  {L.}~\bibnamefont {Zhou}}, \bibinfo {author} {\bibfnamefont {C.~J.}\
  \bibnamefont {Kopas}}, \bibinfo {author} {\bibfnamefont {J.}~\bibnamefont
  {Marshall}}, \bibinfo {author} {\bibfnamefont {J.~Y.}\ \bibnamefont {Mutus}},
  \bibinfo {author} {\bibfnamefont {J.}~\bibnamefont {Slaughter}},\ and\
  \bibinfo {author} {\bibnamefont {et~al.}},\ }\bibfield  {title} {\bibinfo
  {title} {Quasiparticle spectroscopy, transport, and magnetic properties of
  $\mathrm{Nb}$ films used in superconducting transmon qubits},\ }\href
  {https://doi.org/10.1103/physrevapplied.20.024031} {\bibfield  {journal}
  {\bibinfo  {journal} {Physical Review Applied}\ }\textbf {\bibinfo {volume}
  {20}},\ \bibinfo {pages} {024031} (\bibinfo {year} {2023})}\BibitemShut
  {NoStop}%
\bibitem [{\citenamefont {Freitas}\ \emph {et~al.}(2017)\citenamefont
  {Freitas}, \citenamefont {Gonzalez}, \citenamefont {Nascimento},
  \citenamefont {Takeuchi},\ and\ \citenamefont
  {Passamani}}]{Freitas_Gonzalez_Nascimento_Takeuchi_Passamani_2017}%
  \BibitemOpen
  \bibfield  {author} {\bibinfo {author} {\bibfnamefont {T.}~\bibnamefont
  {Freitas}}, \bibinfo {author} {\bibfnamefont {J.}~\bibnamefont {Gonzalez}},
  \bibinfo {author} {\bibfnamefont {V.}~\bibnamefont {Nascimento}}, \bibinfo
  {author} {\bibfnamefont {A.}~\bibnamefont {Takeuchi}},\ and\ \bibinfo
  {author} {\bibfnamefont {E.}~\bibnamefont {Passamani}},\ }\bibfield  {title}
  {\bibinfo {title} {Negative magnetoresistance in sputtered niobium thin films
  grown on silicon substrates},\ }\href
  {https://doi.org/10.1016/j.cap.2017.07.011} {\bibfield  {journal} {\bibinfo
  {journal} {Current Applied Physics}\ }\textbf {\bibinfo {volume} {17}},\
  \bibinfo {pages} {1532–1538} (\bibinfo {year} {2017})}\BibitemShut
  {NoStop}%
\bibitem [{\citenamefont {David~Henry}\ \emph {et~al.}(2014)\citenamefont
  {David~Henry}, \citenamefont {Wolfley}, \citenamefont {Monson}, \citenamefont
  {Clark}, \citenamefont {Shaner},\ and\ \citenamefont {Jarecki}}]{David_2014}%
  \BibitemOpen
  \bibfield  {author} {\bibinfo {author} {\bibfnamefont {M.}~\bibnamefont
  {David~Henry}}, \bibinfo {author} {\bibfnamefont {S.}~\bibnamefont
  {Wolfley}}, \bibinfo {author} {\bibfnamefont {T.}~\bibnamefont {Monson}},
  \bibinfo {author} {\bibfnamefont {B.~G.}\ \bibnamefont {Clark}}, \bibinfo
  {author} {\bibfnamefont {E.}~\bibnamefont {Shaner}},\ and\ \bibinfo {author}
  {\bibfnamefont {R.}~\bibnamefont {Jarecki}},\ }\bibfield  {title} {\bibinfo
  {title} {Stress dependent oxidation of sputtered niobium and effects on
  superconductivity},\ }\href {https://doi.org/10.1063/1.4866554} {\bibfield
  {journal} {\bibinfo  {journal} {Journal of Applied Physics}\ }\textbf
  {\bibinfo {volume} {115}},\ \bibinfo {pages} {083903} (\bibinfo {year}
  {2014})}\BibitemShut {NoStop}%
\bibitem [{\citenamefont {Deutscher}\ and\ \citenamefont
  {de~Gennes}(1969)}]{Deutscher_1969}%
  \BibitemOpen
  \bibfield  {author} {\bibinfo {author} {\bibfnamefont {G.}~\bibnamefont
  {Deutscher}}\ and\ \bibinfo {author} {\bibfnamefont {P.~G.}\ \bibnamefont
  {de~Gennes}},\ }\bibfield  {title} {\bibinfo {title} {Proximity effects.},\
  }\bibfield  {journal} {\bibinfo  {journal} {Superconductivity. Parks, R. D.
  (ed.). New York, Marcel Dekker, Inc., 1969.}\ }\textbf {\bibinfo {volume}
  {1}},\ \href {https://www.osti.gov/biblio/4842932} {} (\bibinfo {year}
  {1969})\BibitemShut {NoStop}%
\bibitem [{\citenamefont {de~Ory}\ \emph {et~al.}(2024)\citenamefont {de~Ory},
  \citenamefont {Rodriguez}, \citenamefont {Magaz}, \citenamefont {Rollano},
  \citenamefont {Granados},\ and\ \citenamefont
  {Gomez}}]{deory2024lowlosshybridnbau}%
  \BibitemOpen
  \bibfield  {author} {\bibinfo {author} {\bibfnamefont {M.~C.}\ \bibnamefont
  {de~Ory}}, \bibinfo {author} {\bibfnamefont {D.}~\bibnamefont {Rodriguez}},
  \bibinfo {author} {\bibfnamefont {M.~T.}\ \bibnamefont {Magaz}}, \bibinfo
  {author} {\bibfnamefont {V.}~\bibnamefont {Rollano}}, \bibinfo {author}
  {\bibfnamefont {D.}~\bibnamefont {Granados}},\ and\ \bibinfo {author}
  {\bibfnamefont {A.}~\bibnamefont {Gomez}},\ }\bibfield  {title} {\bibinfo
  {title} {Low loss hybrid $\mathrm{Nb}$/$\mathrm{Au}$ superconducting
  resonators for quantum circuit applications},\ }\bibfield  {journal}
  {\bibinfo  {journal} {arXiv}\ }\href
  {https://doi.org/10.48550/arXiv.2401.14764} {10.48550/arXiv.2401.14764}
  (\bibinfo {year} {2024})\BibitemShut {NoStop}%
\bibitem [{\citenamefont {Tinkham}(2004)}]{Tinkham_2004a}%
  \BibitemOpen
  \bibfield  {author} {\bibinfo {author} {\bibfnamefont {M.}~\bibnamefont
  {Tinkham}},\ }\href {https://www.worldcat.org/isbn/9780486134727} {\emph
  {\bibinfo {title} {Introduction to Superconductivity}}},\ \bibinfo {edition}
  {2nd}\ ed.\ (\bibinfo  {publisher} {Dover Publications},\ \bibinfo {address}
  {Mineola, New York},\ \bibinfo {year} {2004})\BibitemShut {NoStop}%
\bibitem [{\citenamefont {Il’in}\ \emph {et~al.}(2005)\citenamefont
  {Il’in}, \citenamefont {Siegel}, \citenamefont {Semenov}, \citenamefont
  {Engel},\ and\ \citenamefont
  {Hubers}}]{Il’in_Siegel_Semenov_Engel_Hubers_2005}%
  \BibitemOpen
  \bibfield  {author} {\bibinfo {author} {\bibfnamefont {K.}~\bibnamefont
  {Il’in}}, \bibinfo {author} {\bibfnamefont {M.}~\bibnamefont {Siegel}},
  \bibinfo {author} {\bibfnamefont {A.}~\bibnamefont {Semenov}}, \bibinfo
  {author} {\bibfnamefont {A.}~\bibnamefont {Engel}},\ and\ \bibinfo {author}
  {\bibfnamefont {H.}~\bibnamefont {Hubers}},\ }\bibfield  {title} {\bibinfo
  {title} {Critical current of {N}b and {N}b{N} thin‐film structures: The
  cross‐section dependence},\ }\href {https://doi.org/10.1002/pssc.200460811}
  {\bibfield  {journal} {\bibinfo  {journal} {{P}hysica {S}tatus {S}olidi (c)}\
  }\textbf {\bibinfo {volume} {2}},\ \bibinfo {pages} {1680–1687} (\bibinfo
  {year} {2005})}\BibitemShut {NoStop}%
\bibitem [{\citenamefont {Sun}\ \emph {et~al.}(2017)\citenamefont {Sun},
  \citenamefont {Dai}, \citenamefont {Zhang}, \citenamefont {Luo},
  \citenamefont {Xie}, \citenamefont {Li},\ and\ \citenamefont
  {Lei}}]{Sun_2017}%
  \BibitemOpen
  \bibfield  {author} {\bibinfo {author} {\bibfnamefont {L.}~\bibnamefont
  {Sun}}, \bibinfo {author} {\bibfnamefont {F.}~\bibnamefont {Dai}}, \bibinfo
  {author} {\bibfnamefont {J.}~\bibnamefont {Zhang}}, \bibinfo {author}
  {\bibfnamefont {J.}~\bibnamefont {Luo}}, \bibinfo {author} {\bibfnamefont
  {C.}~\bibnamefont {Xie}}, \bibinfo {author} {\bibfnamefont {J.}~\bibnamefont
  {Li}},\ and\ \bibinfo {author} {\bibfnamefont {H.}~\bibnamefont {Lei}},\
  }\bibfield  {title} {\bibinfo {title} {The electrical resistivities of
  nanostructured aluminium films at low temperatures},\ }\href
  {https://doi.org/10.1088/1361-6463/aa8736} {\bibfield  {journal} {\bibinfo
  {journal} {Journal of Physics D: Applied Physics}\ }\textbf {\bibinfo
  {volume} {50}},\ \bibinfo {pages} {415302} (\bibinfo {year}
  {2017})}\BibitemShut {NoStop}%
\bibitem [{\citenamefont {Janninck}\ and\ \citenamefont
  {Whitmore}(1966)}]{Janninck_Whitmore_1966}%
  \BibitemOpen
  \bibfield  {author} {\bibinfo {author} {\bibfnamefont {R.}~\bibnamefont
  {Janninck}}\ and\ \bibinfo {author} {\bibfnamefont {D.}~\bibnamefont
  {Whitmore}},\ }\bibfield  {title} {\bibinfo {title} {Electrical conductivity
  and thermoelectric power of niobium dioxide},\ }\href
  {https://doi.org/10.1016/0022-3697(66)90094-1} {\bibfield  {journal}
  {\bibinfo  {journal} {Journal of Physics and Chemistry of Solids}\ }\textbf
  {\bibinfo {volume} {27}},\ \bibinfo {pages} {1183–1187} (\bibinfo {year}
  {1966})}\BibitemShut {NoStop}%
\bibitem [{\citenamefont {Krause}\ \emph {et~al.}(2016)\citenamefont {Krause},
  \citenamefont {Afanas'ev}, \citenamefont {Desmaris}, \citenamefont {Meledin},
  \citenamefont {Pavolotsky}, \citenamefont {Belitsky}, \citenamefont
  {Lubenschenko}, \citenamefont {Batrakov}, \citenamefont {Rudziński},\ and\
  \citenamefont {Pippel}}]{Krause_2016}%
  \BibitemOpen
  \bibfield  {author} {\bibinfo {author} {\bibfnamefont {S.}~\bibnamefont
  {Krause}}, \bibinfo {author} {\bibfnamefont {V.}~\bibnamefont {Afanas'ev}},
  \bibinfo {author} {\bibfnamefont {V.}~\bibnamefont {Desmaris}}, \bibinfo
  {author} {\bibfnamefont {D.}~\bibnamefont {Meledin}}, \bibinfo {author}
  {\bibfnamefont {A.}~\bibnamefont {Pavolotsky}}, \bibinfo {author}
  {\bibfnamefont {V.}~\bibnamefont {Belitsky}}, \bibinfo {author}
  {\bibfnamefont {A.}~\bibnamefont {Lubenschenko}}, \bibinfo {author}
  {\bibfnamefont {A.}~\bibnamefont {Batrakov}}, \bibinfo {author}
  {\bibfnamefont {M.}~\bibnamefont {Rudziński}},\ and\ \bibinfo {author}
  {\bibfnamefont {E.}~\bibnamefont {Pippel}},\ }\bibfield  {title} {\bibinfo
  {title} {Ambient temperature growth of mono- and polycrystalline
  $\mathrm{NbN}$ nanofilms and their surface and composition analysis},\ }\href
  {https://doi.org/10.1109/TASC.2016.2529432} {\bibfield  {journal} {\bibinfo
  {journal} {IEEE Transactions on Applied Superconductivity}\ }\textbf
  {\bibinfo {volume} {26}},\ \bibinfo {pages} {1} (\bibinfo {year}
  {2016})}\BibitemShut {NoStop}%
\bibitem [{\citenamefont {Golubev}\ and\ \citenamefont
  {Kuzmin}(2001)}]{Golubev_Kuzmin_2001}%
  \BibitemOpen
  \bibfield  {author} {\bibinfo {author} {\bibfnamefont {D.}~\bibnamefont
  {Golubev}}\ and\ \bibinfo {author} {\bibfnamefont {L.}~\bibnamefont
  {Kuzmin}},\ }\bibfield  {title} {\bibinfo {title} {Nonequilibrium theory of a
  hot-electron bolometer with normal metal-insulator-superconductor tunnel
  junction},\ }\href {https://doi.org/10.1063/1.1351002} {\bibfield  {journal}
  {\bibinfo  {journal} {Journal of Applied Physics}\ }\textbf {\bibinfo
  {volume} {89}},\ \bibinfo {pages} {6464–6472} (\bibinfo {year}
  {2001})}\BibitemShut {NoStop}%
\bibitem [{\citenamefont {Giazotto}\ \emph {et~al.}(2006)\citenamefont
  {Giazotto}, \citenamefont {Heikkilä}, \citenamefont {Luukanen},
  \citenamefont {Savin},\ and\ \citenamefont {Pekola}}]{Pekola_2006}%
  \BibitemOpen
  \bibfield  {author} {\bibinfo {author} {\bibfnamefont {F.}~\bibnamefont
  {Giazotto}}, \bibinfo {author} {\bibfnamefont {T.~T.}\ \bibnamefont
  {Heikkilä}}, \bibinfo {author} {\bibfnamefont {A.}~\bibnamefont {Luukanen}},
  \bibinfo {author} {\bibfnamefont {A.~M.}\ \bibnamefont {Savin}},\ and\
  \bibinfo {author} {\bibfnamefont {J.~P.}\ \bibnamefont {Pekola}},\ }\bibfield
   {title} {\bibinfo {title} {Opportunities for mesoscopics in thermometry and
  refrigeration: Physics and applications},\ }\href
  {https://doi.org/10.1103/revmodphys.78.217} {\bibfield  {journal} {\bibinfo
  {journal} {Reviews of Modern Physics}\ }\textbf {\bibinfo {volume} {78}},\
  \bibinfo {pages} {217–274} (\bibinfo {year} {2006})}\BibitemShut {NoStop}%
\bibitem [{\citenamefont {Karimi}\ \emph {et~al.}(2022)\citenamefont {Karimi},
  \citenamefont {Chang},\ and\ \citenamefont
  {Pekola}}]{Karimi_Chang_Pekola_2022}%
  \BibitemOpen
  \bibfield  {author} {\bibinfo {author} {\bibfnamefont {B.}~\bibnamefont
  {Karimi}}, \bibinfo {author} {\bibfnamefont {Y.-C.}\ \bibnamefont {Chang}},\
  and\ \bibinfo {author} {\bibfnamefont {J.~P.}\ \bibnamefont {Pekola}},\
  }\bibfield  {title} {\bibinfo {title} {Low temperature characteristics of the
  metal–superconductor nis tunneling thermometer},\ }\href
  {https://doi.org/10.1007/s10909-022-02713-z} {\bibfield  {journal} {\bibinfo
  {journal} {Journal of Low Temperature Physics}\ }\textbf {\bibinfo {volume}
  {207}},\ \bibinfo {pages} {220–225} (\bibinfo {year} {2022})}\BibitemShut
  {NoStop}%
\bibitem [{\citenamefont {Zehnder}\ \emph {et~al.}(1999)\citenamefont
  {Zehnder}, \citenamefont {Lerch}, \citenamefont {Zhao}, \citenamefont
  {Nussbaumer}, \citenamefont {Kirk},\ and\ \citenamefont
  {Ott}}]{Zehnder_1999}%
  \BibitemOpen
  \bibfield  {author} {\bibinfo {author} {\bibfnamefont {A.}~\bibnamefont
  {Zehnder}}, \bibinfo {author} {\bibfnamefont {P.}~\bibnamefont {Lerch}},
  \bibinfo {author} {\bibfnamefont {S.~P.}\ \bibnamefont {Zhao}}, \bibinfo
  {author} {\bibfnamefont {T.}~\bibnamefont {Nussbaumer}}, \bibinfo {author}
  {\bibfnamefont {E.~C.}\ \bibnamefont {Kirk}},\ and\ \bibinfo {author}
  {\bibfnamefont {H.~R.}\ \bibnamefont {Ott}},\ }\bibfield  {title} {\bibinfo
  {title} {Proximity effects in
  $\mathrm{Nb}$/$\mathrm{Al}$--$\mathrm{AlO_x}$--$\mathrm{Al}$/$\mathrm{Nb}$
  superconducting tunneling junctions},\ }\href
  {https://doi.org/10.1103/physrevb.59.8875} {\bibfield  {journal} {\bibinfo
  {journal} {Physical Review B}\ }\textbf {\bibinfo {volume} {59}},\ \bibinfo
  {pages} {8875} (\bibinfo {year} {1999})}\BibitemShut {NoStop}%
\bibitem [{\citenamefont {Latorre}\ \emph {et~al.}(2020)\citenamefont
  {Latorre}, \citenamefont {Hofer}, \citenamefont {Rudolph},\ and\
  \citenamefont {Wieczorek}}]{martiSUST2020}%
  \BibitemOpen
  \bibfield  {author} {\bibinfo {author} {\bibfnamefont {M.~G.}\ \bibnamefont
  {Latorre}}, \bibinfo {author} {\bibfnamefont {J.}~\bibnamefont {Hofer}},
  \bibinfo {author} {\bibfnamefont {M.}~\bibnamefont {Rudolph}},\ and\ \bibinfo
  {author} {\bibfnamefont {W.}~\bibnamefont {Wieczorek}},\ }\bibfield  {title}
  {\bibinfo {title} {Chip-based superconducting traps for levitation of
  micrometer-sized particles in the meissner state},\ }\href
  {https://doi.org/10.1088/1361-6668/aba6e1} {\bibfield  {journal} {\bibinfo
  {journal} {Superconductor Science and Technology}\ }\textbf {\bibinfo
  {volume} {33}},\ \bibinfo {pages} {105002} (\bibinfo {year}
  {2020})}\BibitemShut {NoStop}%
\bibitem [{\citenamefont {Schmidt}\ \emph {et~al.}(2024)\citenamefont
  {Schmidt}, \citenamefont {Claessen}, \citenamefont {Higgins}, \citenamefont
  {Hofer}, \citenamefont {Hansen}, \citenamefont {Asenbaum}, \citenamefont
  {Zemlicka}, \citenamefont {Uhl}, \citenamefont {Kleiner}, \citenamefont
  {Gross}, \citenamefont {Huebl}, \citenamefont {Trupke},\ and\ \citenamefont
  {Aspelmeyer}}]{Schmidt2024}%
  \BibitemOpen
  \bibfield  {author} {\bibinfo {author} {\bibfnamefont {P.}~\bibnamefont
  {Schmidt}}, \bibinfo {author} {\bibfnamefont {R.}~\bibnamefont {Claessen}},
  \bibinfo {author} {\bibfnamefont {G.}~\bibnamefont {Higgins}}, \bibinfo
  {author} {\bibfnamefont {J.}~\bibnamefont {Hofer}}, \bibinfo {author}
  {\bibfnamefont {J.~J.}\ \bibnamefont {Hansen}}, \bibinfo {author}
  {\bibfnamefont {P.}~\bibnamefont {Asenbaum}}, \bibinfo {author}
  {\bibfnamefont {M.}~\bibnamefont {Zemlicka}}, \bibinfo {author}
  {\bibfnamefont {K.}~\bibnamefont {Uhl}}, \bibinfo {author} {\bibfnamefont
  {R.}~\bibnamefont {Kleiner}}, \bibinfo {author} {\bibfnamefont
  {R.}~\bibnamefont {Gross}}, \bibinfo {author} {\bibfnamefont
  {H.}~\bibnamefont {Huebl}}, \bibinfo {author} {\bibfnamefont
  {M.}~\bibnamefont {Trupke}},\ and\ \bibinfo {author} {\bibfnamefont
  {M.}~\bibnamefont {Aspelmeyer}},\ }\bibfield  {title} {\bibinfo {title}
  {Remote sensing of a levitated superconductor with a flux-tunable microwave
  cavity},\ }\href {https://doi.org/10.1103/PhysRevApplied.22.014078}
  {\bibfield  {journal} {\bibinfo  {journal} {Phys. Rev. Appl.}\ }\textbf
  {\bibinfo {volume} {22}},\ \bibinfo {pages} {014078} (\bibinfo {year}
  {2024})}\BibitemShut {NoStop}%
\bibitem [{\citenamefont {Chen}\ \emph {et~al.}(2014)\citenamefont {Chen},
  \citenamefont {Neill}, \citenamefont {Roushan}, \citenamefont {Leung},
  \citenamefont {Fang}, \citenamefont {Barends}, \citenamefont {Kelly},
  \citenamefont {Campbell}, \citenamefont {Chen}, \citenamefont {Chiaro},\ and\
  \citenamefont
  {et~al.}}]{Chen_Neill_Roushan_Leung_Fang_Barends_Kelly_Campbell_Chen_Chiaro_etal._2014}%
  \BibitemOpen
  \bibfield  {author} {\bibinfo {author} {\bibfnamefont {Y.}~\bibnamefont
  {Chen}}, \bibinfo {author} {\bibfnamefont {C.}~\bibnamefont {Neill}},
  \bibinfo {author} {\bibfnamefont {P.}~\bibnamefont {Roushan}}, \bibinfo
  {author} {\bibfnamefont {N.}~\bibnamefont {Leung}}, \bibinfo {author}
  {\bibfnamefont {M.}~\bibnamefont {Fang}}, \bibinfo {author} {\bibfnamefont
  {R.}~\bibnamefont {Barends}}, \bibinfo {author} {\bibfnamefont
  {J.}~\bibnamefont {Kelly}}, \bibinfo {author} {\bibfnamefont
  {B.}~\bibnamefont {Campbell}}, \bibinfo {author} {\bibfnamefont
  {Z.}~\bibnamefont {Chen}}, \bibinfo {author} {\bibfnamefont {B.}~\bibnamefont
  {Chiaro}},\ and\ \bibinfo {author} {\bibnamefont {et~al.}},\ }\bibfield
  {title} {\bibinfo {title} {Qubit architecture with high coherence and fast
  tunable coupling},\ }\href {https://doi.org/10.1103/physrevlett.113.220502}
  {\bibfield  {journal} {\bibinfo  {journal} {Physical Review Letters}\
  }\textbf {\bibinfo {volume} {113}},\ \bibinfo {pages} {220502} (\bibinfo
  {year} {2014})}\BibitemShut {NoStop}%
\bibitem [{\citenamefont {Niu}\ \emph {et~al.}(2024)\citenamefont {Niu},
  \citenamefont {Gao}, \citenamefont {He}, \citenamefont {Wang}, \citenamefont
  {Wang},\ and\ \citenamefont {Lin}}]{Niu_Gao_He_Wang_Wang_Lin_2024}%
  \BibitemOpen
  \bibfield  {author} {\bibinfo {author} {\bibfnamefont {Z.}~\bibnamefont
  {Niu}}, \bibinfo {author} {\bibfnamefont {W.}~\bibnamefont {Gao}}, \bibinfo
  {author} {\bibfnamefont {X.}~\bibnamefont {He}}, \bibinfo {author}
  {\bibfnamefont {Y.}~\bibnamefont {Wang}}, \bibinfo {author} {\bibfnamefont
  {Z.}~\bibnamefont {Wang}},\ and\ \bibinfo {author} {\bibfnamefont {Z.-R.}\
  \bibnamefont {Lin}},\ }\bibfield  {title} {\bibinfo {title} {{DC} flux
  crosstalk reduction with dual flux line},\ }\href
  {https://doi.org/10.1063/5.0208859} {\bibfield  {journal} {\bibinfo
  {journal} {Applied Physics Letters}\ }\textbf {\bibinfo {volume} {124}},\
  \bibinfo {pages} {254002} (\bibinfo {year} {2024})}\BibitemShut {NoStop}%
\bibitem [{\citenamefont {Haberkorn}\ \emph {et~al.}(2024)\citenamefont
  {Haberkorn}, \citenamefont {Lee}, \citenamefont {Lee}, \citenamefont {Yun},
  \citenamefont {Verón~Lagger}, \citenamefont {Sirena},\ and\ \citenamefont
  {Kim}}]{Haberkorn_Lee_Lee_Yun_Verón_Lagger_Sirena_Kim_2024}%
  \BibitemOpen
  \bibfield  {author} {\bibinfo {author} {\bibfnamefont {N.}~\bibnamefont
  {Haberkorn}}, \bibinfo {author} {\bibfnamefont {Y.}~\bibnamefont {Lee}},
  \bibinfo {author} {\bibfnamefont {C.}~\bibnamefont {Lee}}, \bibinfo {author}
  {\bibfnamefont {J.}~\bibnamefont {Yun}}, \bibinfo {author} {\bibfnamefont
  {F.}~\bibnamefont {Verón~Lagger}}, \bibinfo {author} {\bibfnamefont
  {M.}~\bibnamefont {Sirena}},\ and\ \bibinfo {author} {\bibfnamefont
  {J.}~\bibnamefont {Kim}},\ }\bibfield  {title} {\bibinfo {title} {Probing the
  quasiparticle relaxation time in ultrathin $\mathrm{NbN}$ microbridges via
  larkin–ovchinnikov instability near tc},\ }\href
  {https://doi.org/10.1021/acsaelm.4c00629} {\bibfield  {journal} {\bibinfo
  {journal} {ACS Applied Electronic Materials}\ }\textbf {\bibinfo {volume}
  {6}},\ \bibinfo {pages} {5077} (\bibinfo {year} {2024})}\BibitemShut
  {NoStop}%
\bibitem [{\citenamefont {Du}\ \emph {et~al.}(2024)\citenamefont {Du},
  \citenamefont {Xu}, \citenamefont {Zhang}, \citenamefont {Li}, \citenamefont
  {Wei}, \citenamefont {Wang}, \citenamefont {Hou}, \citenamefont {Chen},
  \citenamefont {Liu}, \citenamefont {Liu},\ and\ \citenamefont
  {et~al.}}]{Du_Xu_Zhang_Li_Wei_Wang_Hou_Chen_Liu_Liu_et_al._2024}%
  \BibitemOpen
  \bibfield  {author} {\bibinfo {author} {\bibfnamefont {H.}~\bibnamefont
  {Du}}, \bibinfo {author} {\bibfnamefont {Z.}~\bibnamefont {Xu}}, \bibinfo
  {author} {\bibfnamefont {P.}~\bibnamefont {Zhang}}, \bibinfo {author}
  {\bibfnamefont {D.}~\bibnamefont {Li}}, \bibinfo {author} {\bibfnamefont
  {Z.}~\bibnamefont {Wei}}, \bibinfo {author} {\bibfnamefont {Z.}~\bibnamefont
  {Wang}}, \bibinfo {author} {\bibfnamefont {S.}~\bibnamefont {Hou}}, \bibinfo
  {author} {\bibfnamefont {B.}~\bibnamefont {Chen}}, \bibinfo {author}
  {\bibfnamefont {T.}~\bibnamefont {Liu}}, \bibinfo {author} {\bibfnamefont
  {R.}~\bibnamefont {Liu}},\ and\ \bibinfo {author} {\bibnamefont {et~al.}},\
  }\bibfield  {title} {\bibinfo {title} {High-energy electron injection in
  top-gated niobium microbridges for enhanced power efficiency and localized
  control},\ }\href {https://doi.org/10.1063/5.0195254} {\bibfield  {journal}
  {\bibinfo  {journal} {Applied Physics Letters}\ }\textbf {\bibinfo {volume}
  {124}},\ \bibinfo {pages} {112601} (\bibinfo {year} {2024})}\BibitemShut
  {NoStop}%
\bibitem [{\citenamefont {López-Núñez}\ \emph {et~al.}(2023)\citenamefont
  {López-Núñez}, \citenamefont {Montserrat}, \citenamefont {Rius},
  \citenamefont {Bertoldo}, \citenamefont {Torras-Coloma}, \citenamefont
  {Martínez},\ and\ \citenamefont {Forn-Díaz}}]{lopez_2023}%
  \BibitemOpen
  \bibfield  {author} {\bibinfo {author} {\bibfnamefont {D.}~\bibnamefont
  {López-Núñez}}, \bibinfo {author} {\bibfnamefont {Q.~P.}\ \bibnamefont
  {Montserrat}}, \bibinfo {author} {\bibfnamefont {G.}~\bibnamefont {Rius}},
  \bibinfo {author} {\bibfnamefont {E.}~\bibnamefont {Bertoldo}}, \bibinfo
  {author} {\bibfnamefont {A.}~\bibnamefont {Torras-Coloma}}, \bibinfo {author}
  {\bibfnamefont {M.}~\bibnamefont {Martínez}},\ and\ \bibinfo {author}
  {\bibfnamefont {P.}~\bibnamefont {Forn-Díaz}},\ }\bibfield  {title}
  {\bibinfo {title} {Magnetic penetration depth of aluminum thin films},\
  }\bibfield  {journal} {\bibinfo  {journal} {arXiv}\ }\href
  {https://doi.org/10.48550/arXiv.2311.14119} {10.48550/arXiv.2311.14119}
  (\bibinfo {year} {2023})\BibitemShut {NoStop}%
\bibitem [{\citenamefont {Glowacka}\ \emph {et~al.}(2014)\citenamefont
  {Glowacka}, \citenamefont {Goldie}, \citenamefont {Withington}, \citenamefont
  {Muhammad}, \citenamefont {Yassin},\ and\ \citenamefont
  {Tan}}]{Glowacka2014-vq}%
  \BibitemOpen
  \bibfield  {author} {\bibinfo {author} {\bibfnamefont {D.~M.}\ \bibnamefont
  {Glowacka}}, \bibinfo {author} {\bibfnamefont {D.~J.}\ \bibnamefont
  {Goldie}}, \bibinfo {author} {\bibfnamefont {S.}~\bibnamefont {Withington}},
  \bibinfo {author} {\bibfnamefont {H.}~\bibnamefont {Muhammad}}, \bibinfo
  {author} {\bibfnamefont {G.}~\bibnamefont {Yassin}},\ and\ \bibinfo {author}
  {\bibfnamefont {B.~K.}\ \bibnamefont {Tan}},\ }\bibfield  {title} {\bibinfo
  {title} {Development of a $\mathrm{NbN}$ deposition process for
  superconducting quantum sensors},\ }\bibfield  {journal} {\bibinfo  {journal}
  {arXiv}\ }\href {https://doi.org/10.48550/arXiv.1401.2292}
  {10.48550/arXiv.1401.2292} (\bibinfo {year} {2014})\BibitemShut {NoStop}%
\bibitem [{\citenamefont {Salvadori}\ \emph {et~al.}(2004)\citenamefont
  {Salvadori}, \citenamefont {Vaz}, \citenamefont {Farias},\ and\ \citenamefont
  {Cattani}}]{SALVADORI_VAZ_FARIAS_CATTANI_2004}%
  \BibitemOpen
  \bibfield  {author} {\bibinfo {author} {\bibfnamefont {M.~C.}\ \bibnamefont
  {Salvadori}}, \bibinfo {author} {\bibfnamefont {A.~R.}\ \bibnamefont {Vaz}},
  \bibinfo {author} {\bibfnamefont {R.~J.}\ \bibnamefont {Farias}},\ and\
  \bibinfo {author} {\bibfnamefont {M.}~\bibnamefont {Cattani}},\ }\bibfield
  {title} {\bibinfo {title} {Electrical resistivity of nanostructured platinum
  and gold thin films},\ }\href {https://doi.org/10.1142/s0218625x04006086}
  {\bibfield  {journal} {\bibinfo  {journal} {Surface Review and Letters}\
  }\textbf {\bibinfo {volume} {11}},\ \bibinfo {pages} {223–227} (\bibinfo
  {year} {2004})}\BibitemShut {NoStop}%
\bibitem [{\citenamefont {Sambles}\ \emph {et~al.}(1982)\citenamefont
  {Sambles}, \citenamefont {Elsom},\ and\ \citenamefont
  {Jarvis}}]{Sambles_Elsom_Jarvis_1982}%
  \BibitemOpen
  \bibfield  {author} {\bibinfo {author} {\bibfnamefont {J.~R.}\ \bibnamefont
  {Sambles}}, \bibinfo {author} {\bibfnamefont {K.~C.}\ \bibnamefont {Elsom}},\
  and\ \bibinfo {author} {\bibfnamefont {D.~J.}\ \bibnamefont {Jarvis}},\
  }\bibfield  {title} {\bibinfo {title} {The electrical resistivity of gold
  films},\ }\href {https://doi.org/10.1098/rsta.1982.0016} {\bibfield
  {journal} {\bibinfo  {journal} {Philosophical Transactions of the Royal
  Society of London. Series A, Mathematical and Physical Sciences}\ }\textbf
  {\bibinfo {volume} {304}},\ \bibinfo {pages} {365–396} (\bibinfo {year}
  {1982})}\BibitemShut {NoStop}%
\bibitem [{\citenamefont {Joshi}\ \emph {et~al.}(2005)\citenamefont {Joshi},
  \citenamefont {Debnath}, \citenamefont {Aswal}, \citenamefont {Muthe},
  \citenamefont {Senthil~Kumar}, \citenamefont {Gupta},\ and\ \citenamefont
  {Yakhmi}}]{Joshi_2005}%
  \BibitemOpen
  \bibfield  {author} {\bibinfo {author} {\bibfnamefont {N.}~\bibnamefont
  {Joshi}}, \bibinfo {author} {\bibfnamefont {A.}~\bibnamefont {Debnath}},
  \bibinfo {author} {\bibfnamefont {D.}~\bibnamefont {Aswal}}, \bibinfo
  {author} {\bibfnamefont {K.}~\bibnamefont {Muthe}}, \bibinfo {author}
  {\bibfnamefont {M.}~\bibnamefont {Senthil~Kumar}}, \bibinfo {author}
  {\bibfnamefont {S.}~\bibnamefont {Gupta}},\ and\ \bibinfo {author}
  {\bibfnamefont {J.}~\bibnamefont {Yakhmi}},\ }\bibfield  {title} {\bibinfo
  {title} {Morphology and resistivity of $\mathrm{Al}$ thin films grown on
  $\mathrm{Si}$ (111) by molecular beam epitaxy},\ }\href
  {https://doi.org/10.1016/j.vacuum.2005.03.007} {\bibfield  {journal}
  {\bibinfo  {journal} {Vacuum}\ }\textbf {\bibinfo {volume} {79}},\ \bibinfo
  {pages} {178–185} (\bibinfo {year} {2005})}\BibitemShut {NoStop}%
\bibitem [{\citenamefont {Hall}(1968)}]{Hall_1968}%
  \BibitemOpen
  \bibfield  {author} {\bibinfo {author} {\bibfnamefont {L.~A.}\ \bibnamefont
  {Hall}},\ }\href {https://doi.org/10.6028/nbs.tn.365} {\emph {\bibinfo
  {title} {Survey of electrical resistivity measurements on 16 pure metals in
  the temperature range 0 to 273 K}}},\ \bibinfo {type} {technical report}\
  (\bibinfo  {institution} {National Bureau of Standards},\ \bibinfo {year}
  {1968})\BibitemShut {NoStop}%
\bibitem [{\citenamefont {He}\ \emph {et~al.}(2012)\citenamefont {He},
  \citenamefont {Xu}, \citenamefont {Foroughi‐Abari},\ and\ \citenamefont
  {Karpuzov}}]{He_Xu_Foroughi‐Abari_Karpuzov_2012}%
  \BibitemOpen
  \bibfield  {author} {\bibinfo {author} {\bibfnamefont {A.}~\bibnamefont
  {He}}, \bibinfo {author} {\bibfnamefont {S.}~\bibnamefont {Xu}}, \bibinfo
  {author} {\bibfnamefont {A.}~\bibnamefont {Foroughi‐Abari}},\ and\ \bibinfo
  {author} {\bibfnamefont {D.}~\bibnamefont {Karpuzov}},\ }\bibfield  {title}
  {\bibinfo {title} {Depth profiling of $\mathrm{Nb_xO}$/$\mathrm{W}$
  multilayers: Effect of primary ion beam species ($\mathrm{O_2}^+$,
  $\mathrm{Ar}^+$ and $\mathrm{Cs}^+$)},\ }\href
  {https://doi.org/10.1002/sia.4849} {\bibfield  {journal} {\bibinfo  {journal}
  {Surface and Interface Analysis}\ }\textbf {\bibinfo {volume} {44}},\
  \bibinfo {pages} {934–937} (\bibinfo {year} {2012})}\BibitemShut {NoStop}%
\bibitem [{\citenamefont {Komenou}\ \emph {et~al.}(1968)\citenamefont
  {Komenou}, \citenamefont {Yamashita},\ and\ \citenamefont
  {Onodera}}]{Komenou_Yamashita_Onodera_1968}%
  \BibitemOpen
  \bibfield  {author} {\bibinfo {author} {\bibfnamefont {K.}~\bibnamefont
  {Komenou}}, \bibinfo {author} {\bibfnamefont {T.}~\bibnamefont {Yamashita}},\
  and\ \bibinfo {author} {\bibfnamefont {Y.}~\bibnamefont {Onodera}},\
  }\bibfield  {title} {\bibinfo {title} {Energy gap measurement of niobium
  nitride},\ }\href {https://doi.org/10.1016/0375-9601(68)90319-8} {\bibfield
  {journal} {\bibinfo  {journal} {Physics Letters A}\ }\textbf {\bibinfo
  {volume} {28}},\ \bibinfo {pages} {335–336} (\bibinfo {year}
  {1968})}\BibitemShut {NoStop}%
\bibitem [{\citenamefont {Averill}\ \emph {et~al.}(1972)\citenamefont
  {Averill}, \citenamefont {Straus},\ and\ \citenamefont
  {Gregory}}]{Averill_Straus_Gregory_1972}%
  \BibitemOpen
  \bibfield  {author} {\bibinfo {author} {\bibfnamefont {R.~F.}\ \bibnamefont
  {Averill}}, \bibinfo {author} {\bibfnamefont {L.~S.}\ \bibnamefont
  {Straus}},\ and\ \bibinfo {author} {\bibfnamefont {W.~D.}\ \bibnamefont
  {Gregory}},\ }\bibfield  {title} {\bibinfo {title} {Tunneling measurements of
  the superconducting energy gap of bulk polycrystalline indium},\ }\href
  {https://doi.org/10.1063/1.1654043} {\bibfield  {journal} {\bibinfo
  {journal} {Applied Physics Letters}\ }\textbf {\bibinfo {volume} {20}},\
  \bibinfo {pages} {55–56} (\bibinfo {year} {1972})}\BibitemShut {NoStop}%
\bibitem [{\citenamefont {Anderson}\ and\ \citenamefont
  {Ginsberg}(1972)}]{Anderson_1972}%
  \BibitemOpen
  \bibfield  {author} {\bibinfo {author} {\bibfnamefont {R.~A.}\ \bibnamefont
  {Anderson}}\ and\ \bibinfo {author} {\bibfnamefont {D.~M.}\ \bibnamefont
  {Ginsberg}},\ }\bibfield  {title} {\bibinfo {title} {Penetration depth and
  flux creep in thin superconducting indium films},\ }\href
  {https://doi.org/10.1103/PhysRevB.5.4421} {\bibfield  {journal} {\bibinfo
  {journal} {Phys. Rev. B}\ }\textbf {\bibinfo {volume} {5}},\ \bibinfo {pages}
  {4421} (\bibinfo {year} {1972})}\BibitemShut {NoStop}%
\bibitem [{\citenamefont {Maxwell}\ and\ \citenamefont
  {Lutes}(1954)}]{Maxwell_1954}%
  \BibitemOpen
  \bibfield  {author} {\bibinfo {author} {\bibfnamefont {E.}~\bibnamefont
  {Maxwell}}\ and\ \bibinfo {author} {\bibfnamefont {O.~S.}\ \bibnamefont
  {Lutes}},\ }\bibfield  {title} {\bibinfo {title} {Threshold field properties
  of some superconductors},\ }\href {https://doi.org/10.1103/PhysRev.95.333}
  {\bibfield  {journal} {\bibinfo  {journal} {Phys. Rev.}\ }\textbf {\bibinfo
  {volume} {95}},\ \bibinfo {pages} {333} (\bibinfo {year} {1954})}\BibitemShut
  {NoStop}%
\end{thebibliography}%

\end{document}